\documentclass[aps,reprint,twocolumns,pra,showpacs,superscriptaddress,floatfix,notitlepage,nofootinbib]{revtex4-1}
\usepackage{amssymb,amsmath,amsfonts} 
\usepackage{amsthm}
\usepackage{mathtools}
\usepackage{physics}
\usepackage{graphicx,epsfig}

\begin{document}

\title{Stringent Test on Power Spectrum of Quantum Integrable and Chaotic Systems}

\author{A. L. Corps}
  %  \email[Correspondence email address: ]{armando.relano@fis.ucm.es}% Your name
    \affiliation{Departamento de Estructura de la Materia, F\'{i}sica T\'{e}rmica y Electr\'{o}nica, Universidad Complutense de Madrid, Av. Complutense s/n, E-28040 Madrid, Spain}
\author{A. Rela\~{n}o}
\email[Correspondence email address: ]{armando.relano@fis.ucm.es}
\affiliation{Departamento de Estructura de la Materia, F\'{i}sica T\'{e}rmica y Electr\'{o}nica and GISC, Universidad Complutense de Madrid, Av. Complutense s/n, E-28040 Madrid, Spain}

\date{\today} % Leave empty to omit a date

\begin{abstract}
Quantum chaotic and integrable systems are known to exhibit a characteristic $1/f$ and $1/f^{2}$ noise, respectively, in the power spectrum associated to their spectral fluctuations. A recent work  [R. Riser, V. A. Osipov, and E. Kanzieper, \textit{Power Spectrum of Long Eigenlevel Sequences in Quantum Chaotic Systems}, Phys. Rev. Lett. \textbf{118}, 204101 (2017)] calls into question the approximations used to derive these results from random matrix theory. In this paper we show that such approximations do remain valid under almost any circumstances. For the integrable limit, we devise a protocol to exactly recover the original results. As a corollary, we show that the theoretical predictions for other statistics are bound for failure regarding long-range correlations, due to unavoidable spurious effects emerging from the analysis. By means of a rigorous statistical test, we also show that the corrections for the chaotic case introduced in the aforementioned paper require huge statistics to become relevant ---averages over more than $1000$ realizations are mandatory. As an application, we study a paradigmatic model for the crossover from the thermal to the many-body localized phase. We show that our protocol succeeds in describing the crossover. Furthermore, it also succeeds in proving that the Gaussian $\beta$-ensemble fails to account for long-range correlations between the energy levels of this paradigmatic model.

%The theoretical expression for the chaotic case has been derived only up to a good approximation. We perform a rigourous statistical test on this expression to determine the conditions under which this representation, though inexact, can be justified in terms of statistical significance. The theoretical expression for the integrable case, however, is known exactly, but differs from the expression that corresponds to the analogous corrected statistic. We show this difference can be accounted for by means of an analysis of the role the process of unfolding plays in the renormalization. Moreover, we display a explicit transition from the chaotic to the integrable case for a real physical system, in such a way as to apply these theoretical expressions to a practical scenario.
\end{abstract}


\maketitle

\section{Introduction}
Quantum chaos is a relatively young discipline in Physics. Far from establishing a precise, clear definition for the term, the past century saw an identification of the energy level fluctuation properties of a given quantum system and the behavior of its classical analog for large time scales \cite{stockmann}. This resulted in a new picture by which one could understand the underlying classical dynamics of a quantum system by means of carefully analyzing its spectral fluctuations. For quantum systems whose classical analog is integrable, these were shown to be described by Poisson statistics by the pioneering work of Berry and Tabor in the 1970s \cite{berrytabor}, whereas for quantum systems for which the classical limit occurs in a fully chaotic nature, these correspond to the description given by the random matrix theory \cite{rmt}, as was correctly anticipated by Bohigas, Giannoni and Schmit in the 1980s \cite{bgs}. According to this conjecture, level fluctuations in those quantum systems whose classical analog is chaotic will follow one of three classical ensembles, these being Gaussian Orthogonal Ensemble (GOE), Gaussian Unitary Ensemble (GUE), and Gaussian Symplectic Ensemble (GSE). The agreement with one of the previous will strongly depend on the symmetries present in the Hamiltonian under consideration, which preserves the individuality of each system through an analysis based on statistical considerations.

Since then, different approaches have been developed and put to use to offer an accurate description of these spectral fluctuations. We can classify them into two different categories. The first one accounts for short-range correlations between energy levels. The most used statistic within the category is probably the nearest neighbor spacing distribution (NNSD) \cite{stockmann,rmt}, which essentially measures the intensity of level repulsion, to which chaoticity is directly related. However, it is gradually being replaced by the ratio of consecutive level spacings \cite{ratios}, whose main advantage is not requiring the unfolding process of the energy spectrum \cite{misleadingsign}.

The second category deals with long-range correlations between energy levels. Besides traditional $\Sigma_2(L)$ and $\Delta_3(L)$ statistics, describing correlations between energy levels as a function of their distance $L$ \cite{stockmann,rmt}, a characterization in terms of the type of noise present in the spectral fluctuations has acquired an important role \cite{conjetura}. By considering the sequence of energy levels as a discrete time series, it establishes that quantum chaotic systems are characterized by an $1/f$ noise, ubiquitous throughout many complex systems, while for quantum integrable systems this noise becomes $1/f^{2}$. This picture is drawn relying on the power spectrum of the $\delta_n$ statistic. The $1/f$ noise has been experimentally measured in Sinai microwave billiards \cite{billar} as well as in microwave networks \cite{microwave}, whereas an interpolating power-law $1/f^{\alpha}$, with $\alpha\in[1,2)$, was implemented numerically for mixed classical dynamics systems \cite{interpolating}. With the central assumption that for a high enough number of levels in each spectrum this statistic becomes exclusively dominated by two-point correlations, theoretical expressions were derived for the chaotic GOE case in \cite{demo}, as well as for the regular Gaussian Diagonal Ensemble (GDE) case, an ensemble of diagonal matrices with matrix elements coming from a Gaussian distribution which produces Poissonian statistics, and for semiclassical systems, with the help of the random matrix theory. Moreover, results were shown to find solid support in numerical simulations, which expressed an excellent agreement with those. This approach was similarly applied to real physical systems (see \cite{Gomez:11} for a review).

Even though short-range statistics, and especially the ratio of consecutive level spacings, are widely used to get a first idea about the degree of chaos in a quantum system, the analysis of long-range correlations is mandatory to account for an important number of phenomena. The shortest periodic orbits in the semiclassical analogue of a quantum system induce non-universal features which are inaccessible for the NNSD or the ratio of consecutive level spacings. By means of the $\delta_n$ statistic, such non-universal features have been identified in experimental realizations of quantum billiards \cite{billar} and quantum graphs \cite{Dietz17}. A similar effect can appear in spatially extended systems with local Hamiltonians. There exists a characteristic timescale, set by the Thouless time, below which the spectral fluctuations are not well described by random matrix theory. This issue has become the object of current research, relying on the long-range behavior of two-point correlation functions, which are the basis of the $\delta_n$ statistic \cite{Chan18,Bertini18,Kos18}. A typical misleading signature of non-ergodicity is the presence of missing levels ---energy levels which are not detected by the experimental setup--- and mixed symmetries ---fundamental symmetries of the system which are not properly identified, either experimentally or numerically. Both effects are indistinguishable if one just works with short-range statistics. A protocol to properly identify and quantify these effects is based on the original theoretical results for the $\delta_n$ statistic \cite{Molina07}. It has been successfully applied to experimental data of molecular resonances \cite{Mur15}, microwave graphs with violated time reversal invariance \cite{Bialous16} and three-dimensional chaotic microwave cavities \cite{Lawniczak18}, just to quote a few recent papers. 

All these facts reflect the importance of correctly studying long-range spectral statistics. Therefore, the recent publication concerning the conditions under which the theoretical power spectrum of the $\delta_n$ statistic needs to be derived \cite{losenemigos} acquires a broad scope. This work seems to invalidate the method shown in \cite{demo} ---namely, that the power spectrum of a quantum system is solely determined by the spectral form factors associated to it. Indeed, according to \cite{losenemigos}, the power spectrum cannot be merely dominated by two-point correlations: to the contrary, as frequency is increased, this spectrum is increasingly influenced by spectral correlation functions of all orders or, equivalently, the power spectrum keeps record of all orders interactions. The main result given in the work is a parameter-free nonperturbative prediction for the power spectrum for GUE, which is different from the one published in \cite{demo}. Notwithstanding, even more challenging is a corollary of such work, which states that the theoretical result for the integrable GDE is twice the one derived in \cite{demo}. If it is mandatory to apply such results to the usual protocols to obtain the power spectrum of the $\delta_n$ statistic, all the conclusions in papers like \cite{billar,Dietz17,Molina07,Mur15,Bialous16,Lawniczak18} have to be called into question. And an application of the $\delta_n$ statistic to the determination of the Thouless time cannot be accomplished in terms of the rather simple results of \cite{demo}. This means that the cumbersome expressions for GUE published in \cite{losenemigos} have to be used in systems with violated time reversal invariance, and that we still lack equivalent expressions for systems having time reversal invariance.

%Although it is formally correct, we believe it is important to note here that the closed-form expression given in the work is quite cumbersome and mistakes can be easily incurred in. As for the other very often used ensembles, \textit{no} mention of the GOE ensemble is made, and the comment they make on the functional form of the GDE theoretical expression, even though, again, formally correct, establishes a correction \textit{virtually impossible} to obtain, since it would require a perfect unfolding procedure that introduces no spurious correlations. It turns out that these last two ensembles are the most frequently used to analyze experimental data since these correspond to the more frequent symmetries one finds in physical systems, while the GUE ensemble is quite rare in this sense. In conjunction with the previous remarks, this means that from a practical point of view the results devised in \cite{demo} remain to this day the best option in terms of applicability.

In this work we intend to supply a justification for the discrepancy
appearing in the theoretical expression for the GDE case with respect
to \cite{demo} and \cite{losenemigos}. We explain how both equations
are actually correct and justify how one needs to proceed to put a
quantum integrable system power spectrum in correspondence with each
one of them, providing a protocol that enables this purpose. Our main
conclusion is that the requirements for the result given in
\cite{losenemigos} are almost impossible to fulfill; that the
applicability of the original expression, given in \cite{demo},
happens to be much wider, and that the usual protocols to obtain the power spectrum of the $\delta_n$ statistic naturally lead to the original results published in \cite{demo}. Next, we perform a stringent test on the theoretical power spectrum for the GOE case given in \cite{demo} and conclude that, even if not strictly correctly derived for the reasons outlined in \cite{losenemigos}, it suffices to provide an excellent analysis of spectral data, and can be used as an outstanding approximate tool to characterize all kinds of phenomena linked to quantum chaos. Finally, we apply our results to a paradigmatic model for the transition from the ergodic to the many-body localized phase, which has been recently studied in \cite{Buijsman19}. We show that the original theoretical results in \cite{demo} properly describe the ergodic and the many-body localized phases. Furthermore, our stringent test also proves that the Gaussian $\beta$-ensemble \cite{betaensemble}, proposed in \cite{Buijsman19} as a good model for the crossover between these two phases, correctly accounts for short-range correlations between energy levels, but provides poor results for long-range correlations. 

The paper is organized as follows. In Sec. \ref{section2} we give the main definitions and summarize the notation used throughout the text. Sec. \ref{section3} deals with the power spectrum of integrable systems. In Sec. \ref{section4} we perform a stringent numerical test on the spectral fluctuations of fully chaotic systems. In Sec. \ref{section5} we apply the previous results to a system which transits from integrability to chaos. Finally, Sec. \ref{section6} gathers the main conclusions.

\section{Notation conventions and definitions}
\label{section2}

In this section we introduce some notation and definitions that will be constantly used without explicit mention throughout this work.

We consider sets of energy spectra consisting of $N$ levels each, denoted by $E_{i}$, $i\in\{1,2,\dots,N\}$.  Let $\{E_{i}\}_{i=1}^{N}$ be a sequence of energies in ascending order, $E_{i}\geq E_{j}$ whenever $i\geq j$, for each spectrum in the ensemble. The \textit{cumulative level density function} $N(E)\equiv \sum_{j=1}^{N}\Theta(E_{j}-E)$ is separated into a \textit{fluctuating part}, $\widetilde{N}(E)$ and a \textit{smooth part}, $\overline{N}(E)$, which varies continuously with $E$. We make use of the latter for the \textit{unfolding transformation}, by means of which a new sequence of dimensionless levels in ascending order $\{\epsilon_{i}\}_{i=1}^{N}$ is obtained via the mapping $E_{i}\mapsto\epsilon_{i}\equiv \overline{N}(E_{i})$, $\forall i\in\{1,2,\dots, N\}$. These \textit{unfolded energies} can be used to study the statistical properties of different systems as well as different parts of the same spectrum accordingly, and remain valid for any quantum system, regardless of their regularity class.

The \textit{nearest neighbor spacing} is defined to be the consecutive level difference $s_{i}\equiv \epsilon_{i+1}-\epsilon_{i}$, $\forall i\in\{1,2,\dots,N-1\}$. The $\delta_{n}$ \textit{statistic} is then taken to be $\delta_{n}\equiv \sum_{i=1}
^{n}\left(s_{i}-\langle s\rangle\right)=\epsilon_{n+1}-\epsilon_{1}-n$, $\forall n\in\{1,2,\dots,N-1\}$, and represents the deviation of the excitation energy of the $(n+1)$-th unfolded level from its mean value in an equispaced spectrum, $\langle \epsilon_{n}\rangle=n$. If we take $n$ to be a discrete time index, a discrete Fourier transform can be applied to the statistic, thus yielding
\begin{equation}\label{hatdelta}\hat{\delta}_{k}\equiv\mathcal{F}[\delta_{n}]=\frac{1}{\sqrt{N}}\sum_{n=0}^{N-1}\delta_{n}\exp\left(\frac{-2\pi ikn}{N}\right)\in\mathbb{C}.
\end{equation} In Fourier space, we will also denote the \textit{$N-$rescaled frequencies} by $\omega_{k}\equiv 2\pi k/N$, for each $k\in\{1,2,\dots,N-1\}$. Real variables are associated to the subscript $n$, whilst for complex variables in Fourier space this identification occurs with the subscript $k$, although in some instances $\omega_{k}$ will indistinctly play a more useful role. All of them are considered discrete.

The \textit{$\delta_{n}$ power spectrum}, which is the object of interest to us, is then defined to be the square modulus of the Fourier transform $\hat{\delta}_{k}$, that is, $P_{k}^{\delta}\equiv \left|\hat{\delta}_{k}\right|^{2}\in\mathbb{R},$ $\forall k\in\{1,2,\dots,N-1\}$, and allows for the study of the regularity of the system in terms of the density spectral fluctuations associated to it.

The positive integer $M\in\mathbb{N}$ will denote the \textit{number of realizations} over which spectral and ensemble averages, denoted by $\langle\cdot\rangle_{M}$, will be carried. We will write $\langle \cdot \rangle$ whenever the theoretical value of such averages is being referred  to. 

To specify the distribution of an arbitrary random variable $X$, we will denote $X\sim Y$ to compactly mean that $X$ has $Y$ as a probability distribution. Note this has nothing in common with an asymptotic behavior.  \textit{Poisson distributions} are identified with exponential random variables $\textrm{Exp}({\lambda})$ with mean $\lambda^{-1}$, whereas the \textit{normal distribution} with mean $\mu$ and variance $\sigma^{2}$ will be denoted by $\mathcal{N}(\mu,\sigma)$, with $\sigma\equiv\sqrt{\sigma^{2}}$ being the standard deviation associated with the variance $\sigma^{2}$. The probability density and distribution functions of these random variables are denoted following the usual conventions. 

Given a one-variable function $f=f(x)$, we will denote parametric dependence on $y$ by the common expression $f(x;y)$. To lighten up the notation, subscripts will incidentally denote actual variables or parametric dependence, $f_{x}\equiv f(x)$, whenever these are used.

\section{The integrable case}
\label{section3}

In this section we reason on the correct theoretical formula for the power spectrum of integrable quantum systems, we study how the introduction of the unfolding means a modification of such an expected value, and determine the conditions under which the usual spectral analysis needs to be considered.  
\subsection{Derivation of exact theoretical expression}
Consider a spectrum made up of $N+1$ energy levels. As expected, the $\delta_{n}$ statistic is defined $\delta_{n}=\sum_{i=1}^{n}(s_{i}-\langle s_{i}\rangle)$, $\forall n\in\{1,2\dots,N\}$. The mean value $\langle s_{i}\rangle$ is defined as an average over a ensemble of equivalent spectra. That is, if $s_{i}^{(k)}$ denotes the $i$-th spacing in the $k$-th realization, then \begin{equation}\label{unity}
    \langle s_{i}\rangle=\lim_{M\to\infty}\frac{1}{M}\sum_{k=1}^{M}s_{i}^{(k)}=1,\,\,\,\forall i\in\{1,\dots, N\}.
\end{equation}
This mean value is not dependent on the value of $i$. Therefore, we shall denote $\langle s\rangle \equiv \langle s_{i}\rangle$, $\forall i$, where now $\langle s\rangle\equiv \lim_{N\to\infty}\frac{1}{N}\sum_{i=1}^{N}s_{i}$. It is worth remarking that 
\begin{equation}
    \langle s\rangle\neq \frac{1}{N}\sum_{i=1}^{N}s_{i}\equiv \langle s(N)\rangle,
\end{equation}
where $\langle s(N)\rangle$ is a mean sample estimator, while $\langle s\rangle$ is the ensemble mean. From Eq. \eqref{unity} we rewrite \begin{equation}
    \delta_{n}=\sum_{i=1}^{n}(s_{i}-1),\,\,\,n\in\{1,\dots,N\}.
\end{equation}
In integrable systems the nearest neighbor spacings are not intrinsically correlated. This can be quantified by setting $s_{i}=\xi_{i}+1$, where the random variables $\xi_{i}$ are such that $\langle \xi_{i}\rangle=0$ and $\langle \xi_{i}\xi_{j}\rangle=\delta_{ij}$, with $\delta_{ij}$ being the Kronecker delta. It follows that for integrable systems the nearest neighbor spacings must verify the conditions 
\begin{subequations}\begin{equation}\label{conditiona}
   \langle s_{i}\rangle=1,
\end{equation}
\begin{equation}\label{conditionb}
\langle s_{i}s_{j}\rangle=\delta_{ij}+1.
\end{equation}\end{subequations}
The second of the above, Eq. \eqref{conditionb}, determines unequivocally how the nearest neighbor spacings are correlated in systems within this regularity class. Note that the definition of the probability density is not needed for this purpose. Therefore, what follows is valid no matter the probability density, even though we must only apply these results to Poisson statistics. Indeed, all non-integrable systems, whether they are chaotic or not, have nearest neighbor spacings where correlation exists.  

More details on the calculation that follows are given in the Appendix. For integrable systems the (averaged) power spectrum of the $\delta_{n}$ statistic can be written \begin{equation}\label{pk}
    \langle P_{k}^{\delta}\rangle=\frac{1}{N}\sum_{\ell=1}^{N}\sum_{m=1}^{N}\langle\delta_{\ell}\delta_{m}\rangle e^{i\omega_{k}(\ell-m)},
\end{equation}
which applies to any frequency $k$. The correlation factor is given by \begin{equation}\label{corr2}
\langle \delta_{\ell}\delta_{m}\rangle=\min\{\ell,m\}.
\end{equation}
Evaluating Eq. \eqref{pk} affords 
\begin{equation}\label{pk2}
\langle P_{k}^{\delta}\rangle=\frac{1}{N}\sum_{\ell=1}^{N}\sum_{m=1}^{N}\min\{\ell, m\}e^{i\omega_{k}(\ell-m)}.
\end{equation}
Performing the double sum over both $\ell$ and $m$ yields 
\begin{equation}\label{power2}
\langle P_{k}^{\delta}\rangle=\frac{1}{2\sin^{2}\left(\frac{\pi k}{N}\right)}+\order{\frac{1}{N}}.
\end{equation}
Note that the expression given in \cite{losenemigos} corresponds to the leading order of such integrable power spectrum. This is the only possible contribution only when $N\to\infty$. However, it is \textit{not correct} that the theoretical power spectrum for integrable systems can be described by such a leading order term by itself for an arbitrary value of $N$, as has been suggested. The $1/f^{2}$ noise must be obtained for $k\ll N$ and $N\gg 1$ accordingly, since the neglected terms also do depend on $k$.  It is under these circumstances that one would obtain the Maclaurin representation given by the inverse power-law \begin{equation}
    \langle P_{k}^{\delta}\rangle\simeq\frac{N^{2}}{2\pi^{2}k^{2}}\propto \frac{1}{k^{2}}.
\end{equation}

\subsection{The role of the unfolding procedure}
We now move on to examine the consequences of the process of unfolding. Even though the previous calculation is formally correct, the hypothesis under which it has been derived are very strict in practical terms. The main reason for this is that the sequence of spacings is obtained after unfolding a set of levels characterized by a density of states that may very well be unknown. More importantly, Eq. \eqref{power2} requires Eq. \eqref{conditionb} to be strictly verified to hold. This condition implies (see Appendix for details) $\langle \delta_{\ell}^{2}\rangle=\ell$. Because every spectrum needs to be unfolded to generate the $\delta_{n}$ statistic, the following questions deserve to be asked: \textit{Does the unfolding procedure \emph{really} guarantee the complete verification of Eq. \eqref{conditionb}? Or, on the contrary, does it introduce some sort of spurious correlation in the statistical analysis? }

For the sake of simplicity, let us assume for a moment the smooth part of the states probability density, $\overline{\rho}(E)$, to be exactly known. Suppose as well that it is normalized such that \begin{equation}
    \int_{-\infty}^{\infty}\mathrm{d}E\,\overline{\rho}(E)=N,
\end{equation}where $N$ is the total number of levels in the spectrum. This is the case for the GDE, for which one has $\overline{\rho}(E)=1/\sqrt{2\pi}\exp{-x^{2}/2}$. This is also the case where spurious effects are expected the least \cite{misleadingsign}. By definition, one has 
\begin{equation}
    \epsilon_{1}=\overline{N}(E_{1})\equiv \int_{-\infty}^{E_{1}}\mathrm{d}E\,\overline{\rho}(E)
\end{equation} and 
\begin{equation}
    \epsilon_{N+1}=\overline{N}(E_{N+1})\equiv \int_{-\infty}^{E_{N+1}}\mathrm{d}E\,\overline{\rho}(E).
\end{equation}
As a consequence, one clearly has $[\epsilon_{1}]_{\min}=0$ as well as $[\epsilon_{N+1}]_{\max}=N+1$. This argument can be generalized to assert that the minimum value of $\epsilon_{i}=\overline{N}(E_{i})$ and the maximum value of $\epsilon_{j}=\overline{N}(E_{j})$ are, respectively, $[\epsilon_{i}]_{\min}=0$ and $[\epsilon_{j}]_{\max}=N+1$, $\forall i,j\in\{1,\dots,N+1\},$ where the equalities are strictly verified. Equivalently, one has $\epsilon_{j}\in(0,N+1)$, where not all possible values are equally probable. Therefore, on calculating the $\delta_{n}$ statistic one finds, for $n=N$, $\delta_{N}=\epsilon_{N+1}-\epsilon_{1}-N.$
The maximum value of the above equation is then simply \begin{equation}\label{maxdelta}[\delta_{N}]_{\max}=[\epsilon_{N+1}]_{\max}-[\epsilon_{1}]_{\min}-N=N+1-N=1.\end{equation}
The result in Eq. \eqref{maxdelta} is in absolute contradiction with Eq. \eqref{conditionb} since, were they to hold, $\langle \delta_{N}^{2}\rangle=N$, and $\delta_{N}$ would be a random variable with mean $\mu=0$ and variance $\sigma^{2}=N$. Observe that the statistic cannot be described by a random walk when the unfolding has been performed analytically on the original energy levels. Hence, we conclude that \textit{the usual way to proceed with the unfolding process introduces spurious correlations in integrable systems. The relevance this bears on quantum chaotic systems is not so vital, because these are already intrinsically correlated. }

A convenient way to emulate these effects is to consider that, as a consequence of the spurious correlations, the unfolding procedure acts on the sequence of spacings so as to 
%As a consequence, the statistic is not a random walk when the unfolding has been performed analitically on the original energy levels. For if it were so, then $\langle \delta_{N}^{2}\rangle=N$, which cannot be satisfied. If one performs the same calculation using the reunfolded spacings instead, this reads for $n=N$ 
%\begin{equation}\begin{split}
%\delta_{N}=\sum_{i=1}^{N}(s_{i}-1)=N\langle s\rangle_{M}-N=N-N=0. 
%\end{split}\end{equation}
 \begin{equation}\label{normalization}
\frac{1}{N}\sum_{i=1}^{N}s_{i}=1,
\end{equation} where the above equality is \textit{exact}. In general, for a given set of spacings $\{s_{i}\}_{i=1}^{N}$ Eq. \eqref{normalization} is not strictly verified, but approximately.

A simple way to account for this effect is to redefine the nearest neighbor spacing to be \begin{equation}\label{scorr}
\widetilde{s}_{i}\equiv \frac{s_{i}}{\langle s_{i}\rangle}=\frac{Ns_{i}}{\sum_{k=1}^{N}s_{k}},\,\,\forall i\in\{1,\dots,N\}.
\end{equation}
The set of spacings $\{\widetilde{s}_{i}\}_{i=1}^{N}$ defined by Eq. \eqref{scorr} will be hereinafter called \textit{reunfolded spacings}. 
%Let us concentrate on the following observation. Whenever an exactly calculated, analytic function is used as the smooth part of the density of states, $\overline{N}$, one has that the minimum value of $\epsilon_{i}=\overline{N}(E_{i})$ and the maximum value of $\epsilon_{j}=\overline{N}(E_{j})$ are, respectively, $[\epsilon_{i}]_{\min}=0$ and $[\epsilon_{j}]_{\max}=N+1$, $\forall i,j\in\{1,\dots,N+1\},$ where the equalities are strictly verified. Equivalently, one has $\epsilon_{j}\in(0,N+1)$, where not all posible values are equiprobable. In particular, one clarly has $[\epsilon_{1}]_{\min}=0$ as well as $[\epsilon_{N+1}]_{\max}=N+1$. Therefore, on calculating the $\delta_{n}$ statistic one finds for $n=N$ \begin{equation}
  %  \delta_{N}=\epsilon_{N+1}-\epsilon_{1}-N.\end{equation}
%The maximum value of the above equation is then simply $[\delta_{N}]_{\max}=[\epsilon_{N+1}]_{\max}-[\epsilon_{1}]_{\min}-N=N+1-N=1.$ 
%As a consequence, the statistic is not a random walk when the unfolding has been performed analitically on the original energy levels. For if it were so, then $\langle \delta_{N}^{2}\rangle=N$, which cannot be satisfied. 
If one performs an analogous calculation using the reunfolded spacings instead, this reads for $n=N$ 
\begin{equation}\begin{split}
\delta_{N}=\sum_{i=1}^{N}(s_{i}-1)=N\langle s\rangle_{M}-N=N-N=0. 
\end{split}\end{equation}
This result is much closer to the property $[\delta_{n}]_{\max}=1$, which has been understood to be a spurious consequence of the unfolding procedure, than it is to the ideal case where $\delta_{N}$ is a random variable with mean $\mu=0$ and variance $\sigma^{2}=N$.

An analogous statistic to the $\delta_{n}$ statistic can be naturally defined as that arising from the new sequence of spacings by the equivalence \begin{equation}\label{deltacorr}
\widetilde{\delta}_{n}\equiv \sum_{i=1}^{N}(\widetilde{s}_{i}-1), \,\,\,\forall n\in\{1,\dots,N\}.
\end{equation}
We will refer to this statistic as the \textit{corrected $\delta_{n}$}. Associated with it is the new \textit{corrected power spectrum}\begin{equation}\label{pcorrk}
\langle \widetilde{P}_{k}^{{\delta}}\rangle\equiv\langle {P}_{k}^{\widetilde{\delta}}\rangle=\frac{1}{N}\sum_{\ell=1}^{N}\sum_{m=1}^{N}\langle \widetilde{\delta}_{\ell}\widetilde{\delta}_{m}\rangle e^{i\omega_{k}(\ell-m)},
\end{equation}
which, after expansion, becomes
\begin{equation}
  \label{pcorrk2}
  \begin{split}
    \langle \widetilde{P}_{k}^{\delta}\rangle &= \frac{1}{N}\sum_{\ell=1}^{N}\sum_{m=1}^{N}\left(\frac{N}{N+1}\min\{\ell,m\} - \right. \\
    &\left. -\frac{\ell m}{N+1}\right)e^{i\omega_{k}(\ell-m)}, 
  \end{split}
  \end{equation}
where the correlation factor is now given by 
\begin{equation}\label{corr4}
\langle \widetilde{\delta}_{\ell}\widetilde{\delta}_{m}\rangle=\frac{N}{N+1}\min\{\ell,m\}-\frac{\ell m}{N+1}.
\end{equation}
Performing the corresponding double sum now yields 
\begin{equation}\label{power4}
\langle \widetilde{P}_{k}^{\delta}\rangle=\frac{1}{4\sin^{2}\left(\frac{\pi k}{N}\right)}+\order{\frac{1}{N}},
\end{equation}
which differs from Eq. \eqref{power2} by a factor of $2$. Clearly, one has $\langle P_{k}^{\delta}\rangle=2\langle \widetilde{P}_{k}^{\delta}\rangle$. The leading order of Eq. \eqref{power4} corresponds with the theoretical power spectrum for the integrable case derived in \cite{demo}. In the domain defined by the conditions $k\ll N$ and $N\gg1$, one analogously has for the Maclaurin representation of Eq. \eqref{power4} the known inverse square power-law
\begin{equation}
    \langle \widetilde{P}_{k}^{\delta}\rangle\simeq \frac{N^{2}}{4\pi^{2}k^{2}}\propto\frac{1}{k^{2}}.
\end{equation}
The calculation is shown at length in the Appendix.

\subsection{Numerical evidence}
To illustrate numerically the above results, we proceed as follows. For the following scenarios, we must keep in mind that we are interested in the best possible unfolding, that is, the one that introduces the least possible noise in our analysis. This requires making use of an analytical exact formula obtained for the cumulative density of states, $\overline{N}$. Since the starting probability distribution is a standard normal and we need it to be normalized to the number of levels participating in the unfolding process, we calculate \begin{equation}\label{accnormal}\begin{split}
    \overline{N}(E)&=N\int_{-\infty}^{E}\mathrm{d}x\,\frac{1}{\sqrt{2\pi}}\exp\left(\frac{-x^{2}}{{2}}\right)\\ &=\frac{N}{2}\left[1-\erf\left(\frac{-E}{\sqrt{2}}\right)\right],
\end{split}\end{equation}
where $\erf(\cdot)$ is Gauss error function. Eq. \eqref{accnormal} is the exact formula employed to find the unfolded levels in this part of the work.  We simultaneously consider two different processes: 
%\hrule
%\vspace{0.1in}

\fbox{\textbf{I}} Generate a vector $A$ of dimension $N=10^{9}$ whose entries are numbers coming from a  standard normal distribution, that is, select a certain $A\in\mathcal{M}_{1\times N}\left(\mathcal{N}(0,1)\right)$ whose entries, after reordering in ascending order if necessary, will be taken as the initial levels $\{E_{i}\}_{i=1}^{N}=\{(A)_{1,i}\}_{i=1}^{N}$. Next, the sequence is unfolded to obtain the list of levels in ascending order $\{\epsilon_{i}\}_{i=1}^{N}=\{\overline{N}(E_{i})\}_{i=1}^{N}$. \textit{Note the unfolding is only performed once.} Fix $M=10^{6}$. Partition the set of unfolded levels into $M$ sets of energies consisting of $N/M=10^{3}$ levels each, that is, construct explicitly the $M$ sets of levels $\left\{\{\epsilon_{i+(j-1)N/M}\}_{i=1}^{N/M}\right\}_{j=1}^{M}$. We expect the spurious effects introduced by the unfolding process to be less important for this case, the reason being that those should have an effect on large distance correlations. Because the unfolded spectrum has been split into spectra with relative length $(N/M)/N=1/M=10^{-6}$, it stands to reason to expect them not to play such an important role. Note that these spurious effects should manifest primarily on the leftmost and rightmost eigenlevels in the spectrum for each realization, and this process explicitly assigns these edges a slight weight over the total ensemble under consideration. Obtain the averaged power spectrum $\langle P_{k}^{\delta}\rangle_{M}$. Conclude it is accurately described by Eq. \eqref{power2}. Construct the random variable $ \delta_{n}^{2}$ and find $\langle \delta_{n}^{2}\rangle_{M}$. Compare the result with the expected value Eq. \eqref{corr2} with $\ell=m\equiv n$, that is, 
\begin{equation}\label{corr2n}
\eval{\langle\delta_{\ell}\delta_{m}\rangle}_{\ell=m=n}=\langle \delta_{n}^{2}\rangle=n.
\end{equation} Observe this second-order moment is expected to behave linearly and to grow unboundedly as $n$ grows. 
%\hrule
%\vspace{0.1in}

\fbox{\textbf{II}} Generate a vector $A$ of dimension $N=10^{3}$ whose entries are numbers coming from a standard normal distribution, that is, select a certain $A\in\mathcal{M}_{1\times N}\left(\mathcal{N}(0,1)\right)$ whose entries, after reordering in ascending order if necessary, will be taken as the initial levels $\{E_{i}\}_{i=1}^{N}=\{(A)_{1,i}\}_{i=1}^{N}$. Proceed as in \textbf{I.}, but \textit{do not partition the set of energies}. Repeat and average over $M=10^{6}$ realizations. \textit{It is noteworthy that the unfolding procedure is considered exactly $M$ times now}. This means \textit{all} $M$ spectra and the eigenlevels located at their leftmost and rightmost positions will suffer from spurious correlations implicitly introduced by the unfolding process, which will be present throughout the whole ensemble. Find the averaged power spectrum $\langle P_{k}^{\delta}\rangle_{M}$. Conclude it is governed by Eq. \eqref{power4}. Construct the random variable $ \delta_{n}^{2}$ and find $\langle \delta_{n}^{2}\rangle_{M}$. Compare the result with the expected value Eq. \eqref{corr4} with $\ell=m\equiv n$, that is, 
\begin{equation}\label{corr4n}
\eval{\langle\widetilde{\delta}_{\ell}\widetilde{\delta}_{m}\rangle}_{\ell=m=n}=\langle \widetilde{\delta}_{n}^{2}\rangle=\frac{nN}{N+1}-\frac{n^{2}}{N+1}.
\end{equation} Observe this second-order moment is expected to be parabolic and to reach its maximum value when $n=N/2$. 
%\hrule
%\vspace{0.1in}

\begin{figure}[h]
\centering
\includegraphics[width=0.54\textwidth]{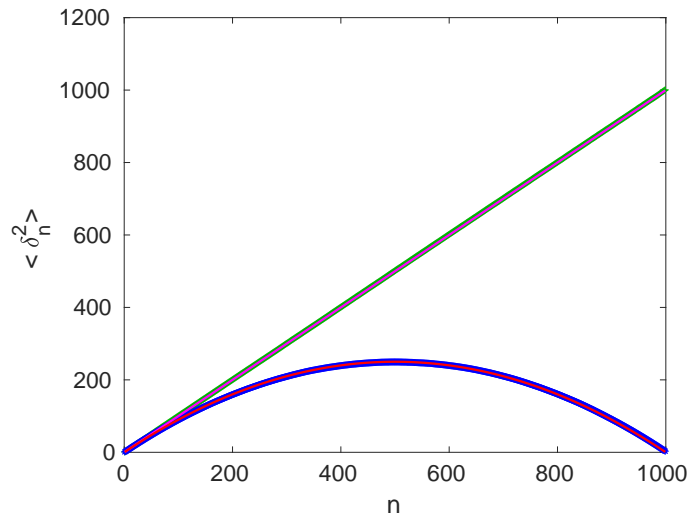}
\caption{(Color online) Theoretical $\langle \delta_{n}^{2}\rangle$ for GDE Eqs. \eqref{corr2n} (top, magenta solid line) and \eqref{corr4n} (bottom, red solid line)  compared to numerical averages consisting of $10^{6}$ spectra with $N=10^{3}$ levels each obtained following processes \textbf{I} (top, green solid points) and \textbf{II} (bottom, blue solid points). }
\label{deltacuadrado}
\end{figure}

\begin{figure}[h]
\centering
\includegraphics[width=0.52\textwidth]{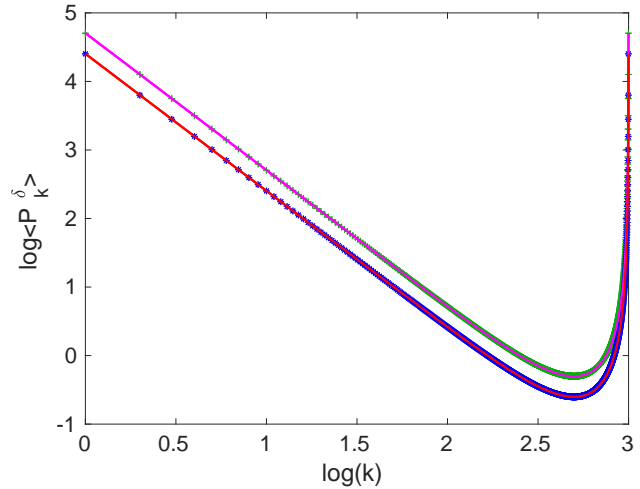}
\caption{(Color online) Theoretical $\langle P_{k}^{\delta}\rangle$ for GDE Eqs. \eqref{power2} and \eqref{power4} (top and bottom solid lines) compared to numerical averages consisting of $10^{6}$ spectra with $N=10^{3}$ levels each obtained following processes \textbf{I} (+ symbols) and \textbf{II} (* symbols).}
\label{GDE}
\end{figure}

The previous steps can be performed algorithmically to obtain the desired results, which are, in essence, the mean $\delta_{n}^{2}$ over the total ensemble and the averaged integrable power spectrum for each different process above detailed. These are shown in Figs. \ref{deltacuadrado} and \ref{GDE}.

In Fig. \ref{deltacuadrado}, we contrast the simulated data $\langle\delta_{n}^{2}\rangle_{M}^{i}$, $i\in\{\mathbf{I},\mathbf{II}\}$ for the processes \textbf{I} and \textbf{II} with the theoretical expected curves. We observe the outstanding agreement with which $\langle\delta_{n}^{2}\rangle_{M}^{\mathbf{I}}$ is put in correspondence with Eq. \eqref{corr2n}, while this is analogously true for $\langle\delta_{n}^{2}\rangle_{M}^{\mathbf{II}}$ and Eq. \eqref{corr4n}, as we had anticipated.
Fig. \ref{deltacuadrado} allows to conclude that, indeed, the nature of spurious correlations can be greatly modified by performing the previous steps, and this relies, basically, on altering the region of the spectrum where the unfolding is permitted to add information that was initially not there. 

In Fig. \ref{GDE}, we confirm the results just recently discussed. As one would expect, different correlation functions are associated with different power spectra, and we have managed to make $\langle P_{k}^{\delta}\rangle_{M}^{\mathbf{I}}$ coincide with the results predicted by Eq. \eqref{power2}, as explained in \textbf{I}, whereas $\langle P_{k}^{\delta}\rangle_{M}^{\mathbf{II}}$ agrees with Eq. \eqref{power4}, according to \textbf{II}. In both cases we observe how the numerical data are equally outstandingly well described by the theoretical curve to which they are related. 

As has been shown, in spite of the fact that the result given in \cite{losenemigos} is, by all means, correct, the unfolding procedure very often followed introduces spurious correlations that act on the unfolded sequence such that they are described by the theoretical expression in \cite{demo}. We note that, for any given spectrum coming from a calculation or an experiment, it is very frequent to proceed with the statistical analysis \textit{without partitioning} our sample of eigenlevels. Our results prove that, in all those cases, one needs to take the results in \cite{demo} into consideration to perform a successful investigation. 

We would like to end this section after a relevant remark. The role the unfolding procedure plays in the study of quantum integrable systems, namely, the introduction of spurious correlations that manifest throughout the whole spectrum, has universal implications that apply to other long-range spectral statistics that are usually jointly considered to make a decision about the regularity of a quantum system. For the \textit{number variance} \begin{equation}
    \Sigma^{2}(L)\equiv \langle (n(E,L)-L)^{2}\rangle,
\end{equation} where $L\gg1$ is the size of an interval in $\mathbb{R}$ and $n(E,L)$ denotes the number of levels in such an interval at energy $E$, the theoretical machinery predicts a linear expected value for the integrable case \begin{equation}\label{sigma2th}
\Sigma^{2}_{\textrm{Int}}(L)=L.
\end{equation} Similarly, long-range correlations of level spacings can also be characterized by means of the Dyson-Mehta \textit{spectral rigidity}, 
\begin{equation}
    \Delta_{3}(E_{0},L)\equiv \frac{1}{L}\min_{\alpha,\beta\in\mathbb{R}}\int_{E_{0}}^{E_{0}+L}\mathrm{d}E\,\left[N(E)-\alpha E-\beta\right]^{2},
\end{equation} for which the integrable regime is governed by the likewise linear expression
\begin{equation}\label{delta3th}
\Delta_{3}^{\textrm{Int}}(E_{0},L)=\frac{L}{15}.
\end{equation}
As can be seen in Eqs. \eqref{sigma2th} and \eqref{delta3th}, these spectral statistics seem to share a commonality with the $\delta_{n}^{2}$ in that the regular regime takes on a linear form as well. This is due to the mathematical hypothesis under which these results have been derived, consisting of completely uncorrelated level spacings (see, preferably, \cite{rmt}). This was to be expected. However, careful accommodation of the spurious correlations, trademark of the unfolding process, implies that this behavior will suffer a modification that goes in the same direction as that of Eq. \eqref{corr4n}, which means the linear structure will be severely influenced by correlations that should in principle not exist in an integrable quantum system, and deviations will, too, occur for both $\Sigma^{2}_{\textrm{Int}}(L)$ and $\Delta_{3}^{\textrm{Int}}(E_{0},L)$, meaning these will no longer be described by such a simple linear functional form. This has already been justified for the $\delta_{n}^{2}$ statistic through the realization that level spacings cease to be uncorrelated as a consequence of the unfolding transformation. It directly follows that, when uncorrelation is also violated for the other two statistics, the same conclusion must be reached. 

Before ending this section, it is worth noting that similar qualitative results, regarding $\Sigma_2(L)$ and $\Delta_3(L)$ statistics, were anticipated in \cite{drozdz}.

\section{The chaotic case}
\label{section4}

The main purpose of this section is to show that the theoretical
expression for the averaged power spectrum $\langle
P_{k}^{\delta}\rangle$ derived in \cite{demo}, making use of spectral
form factors under the hypothesis that two-point correlations dominate
the dynamics, is sufficient to obtain a correct representation of the
behaviour of the system. Indeed, as suggested in \cite{losenemigos},
the procedure to follow if one wishes to find the correct theoretical
expression needs to take into account the reality that the power
spectrum is increasingly influenced by spectral correlation functions
of all orders. Notwithstanding, such an exact formula has not been
deduced as of yet, and in said work no explicit mention of the
structure this expression should have is made. Therefore, it is our
goal to show here that, even if the initial hypothesis that two-point
correlations can be used to describe the power spectrum should not
hold, the numerical agreement is remarkably good, thus justifying its
use to study the chaotic nature of a quantum system.

\subsection{Influence of the number of realizations}

\begin{figure}[h]
\centering
\includegraphics[width=0.52\textwidth]{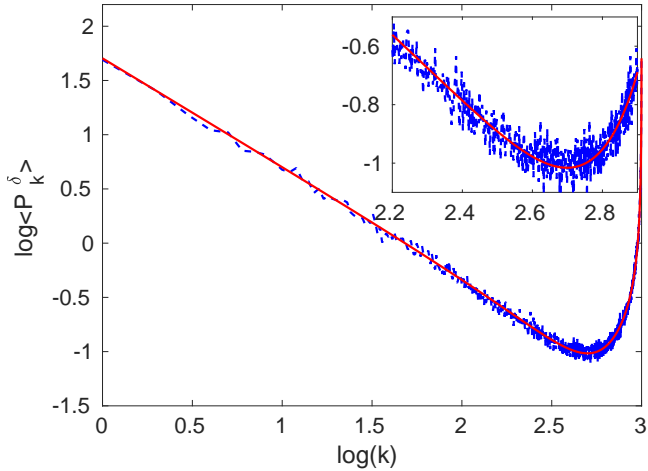}
\caption{(Color online) Theoretical power spectrum $\langle P_{k}^{\delta}\rangle$ of the $\delta_{n}$ function for GOE (solid line), compared to numerical averages consisting of $10^{2}$ spectra with $N=10^{3}$ levels each (dashed line).}
\label{GOE102}
\end{figure}

\begin{figure}[h]
\centering
\includegraphics[width=0.52\textwidth]{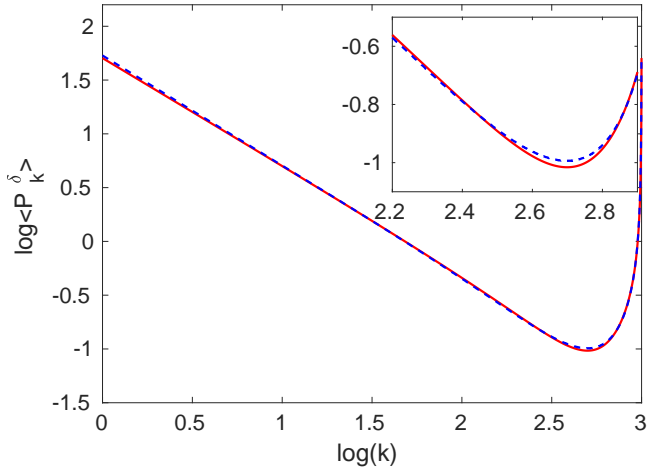}
\caption{(Color online) Theoretical power spectrum $\langle P_{k}^{\delta}\rangle$ of the $\delta_{n}$ function for GOE (solid line), compared to numerical averages consisting of $10^{6}$ spectra with $N=10^{3}$ levels each (dashed line).}
\label{GOE106}
\end{figure}

Explicitly, the two-point correlation theoretical expression for quantum chaotic systems, within the context of Random Matrix Ensembles (RME), takes on the free-parameter form 
\begin{equation}\label{powergoe}
\begin{split}
    \langle P_{k}^{\delta}\rangle&=\frac{N^{2}}{4\pi^{2}}\left[\frac{K\left(\frac{k}{N}\right)-1}{k^{2}}+\frac{K\left(1-\frac{k}{N}\right)-1}{(N-k)^{2}}\right]\\& +\frac{1}{4\sin^{2}\left(\frac{\pi k}{N}\right)}-\frac{1}{12},
    \end{split}
\end{equation}
where $k\in\{1,2,\dots,N-1\}$, $N$ denotes the size of each spectrum in the ensemble over which the average has been performed, and the spectral form factor, $K$, for the GOE case admits the representation \begin{equation}
    K(\tau)=\begin{cases}
{\displaystyle 2\tau-\tau\log(1+2\tau)},\,\,\,&\tau\leq1\\[0.2cm]
{\displaystyle 2-\tau\log\left(\frac{2\tau+1}{2\tau-1}\right)},\,\,\,&\tau\geq1
\end{cases}
\end{equation}
which will be put in comparison with two sets of numerical data. We will consider simultaneously an ensemble of $M=10^{2}$ and $M=10^{6}$ GOE matrices of dimension $N=10^{3}$ each. The corresponding power spectra for the $\delta_{n}$ statistic is then calculated as follows. We first evaluate the smooth cumulative density function $\overline{N}$ at the energy levels $\{E_{i}\}_{i=1}^{N}$. We invoke the properties of the GOE and make use of its exact free-parameter cumulative density obtained, for instance, in \cite{stockmann}, that is,
\begin{equation}
  \begin{split}
\overline{N}(E)&= \frac{E}{2\pi}\sqrt{2N-E^{2}}+ \\ &+\frac{N}{\pi}\arctan\left(\frac{E}{\sqrt{2N-E^{2}}}\right)+ \frac{N}{2},\,\,\, |E|\leq\sqrt{2N},
  \end{split}
  \end{equation}
$\overline{N}(E)=0$, if $E<\sqrt{2N}$, and $\overline{N}(E)=N$, if $E>\sqrt{2N}$, in such a way that the new unfolded levels are uniquely determined. This corresponds to the so-called Wigner's semicircle law. We note that the first and last levels of the original sequence of energies have been eliminated to avoid dealing with spurious effects arising from it. It is worth observing that the sequence $\{\epsilon_{i}\}_{i=1}^{N}$ is in this case perfectly calculated: any attempt to find it by fitting to any polynomial would result in a noisy set of unfolded levels, and this is true independently of the adequacy of the polynomial.  We then construct the nearest neighbor spacings as $s_{i}\equiv \epsilon_{i+1}-\epsilon_{i}$, $\forall i\in\{1,2,\dots,N-1\}$, from which the expression for the $\delta_{n}$ statistic as well as its power spectrum $P_{k}^{\delta}$ is a straightforward matter. In Figs. \ref{GOE102} and \ref{GOE106}, we display the parameter-free prediction with the results of numerical simulations for these sets of spectra. Note that the simulated points have been represented with a dashed line for viewing purposes only.

Fig. \ref{GOE102} clearly establishes the impossibility of ruling out the disagreement between theoretical and simulated points because of the noisy nature of the spectral fluctuations, whilst we observe a discrepancy in the second case, Fig. \ref{GOE106}, for high-lying frequencies, $2.5\lesssim \log(k)\lesssim 2.8$, as well as for low-lying ones, $0\lesssim \log(k)\lesssim0.5$. This serves the initial purpose of suggesting that as the number of realizations involved in the average increases, the accordance of the numerical data and the theoretical curve will decrease. The same conclusion was reached in \cite{losenemigos}.

\begin{figure}[h]
\centering
\includegraphics[width=0.52\textwidth]{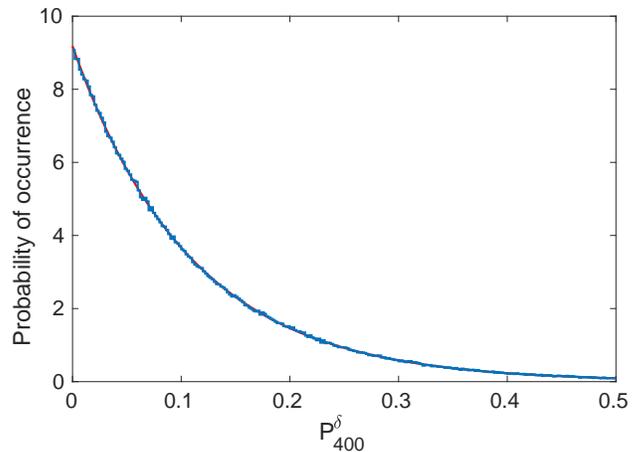}
\caption{(Color online) Probability density for the random variable $ P_{400}^{\delta}$. Numerical simulations carried with $M=10^{6}$ spectra of $N=10^{3}$ levels each. }
\label{histogram}
\end{figure}

In Fig. \ref{histogram} we show the histogram representing the probability density of the random variable $ P_{k}^{\delta}$ for the frequency $k=400$ plotted against a Poisson probability density with mean $ P_{400}^{\delta}.$ The agreement is remarkable, so much so that we have only been able to clearly show one of said two curves. The behaviour is analogous for any other frequency $k$, so that in general the random variable $P_{k}^{\delta}$ for quantum chaotic systems is distributed as \begin{equation}
     P_{k}^{\delta}\sim \textrm{Exp}\left(\frac{1}{ P_{k}^{\delta}}\right).
\end{equation} 
One quickly realizes that since for the chaotic $\langle P_{k}^{\delta}\rangle$ the statistical mode and the mean value are reasonably afar from each other, a single realization will not be representative of the theoretical distribution. This implies that in order to obtain statistically significant results one must go on to consider averages over many realizations. 

\subsection{Significance test on two-point power spectrum result}

We wish to obtain an upper bound for the number of realizations for which the numerical discrepancy cannot be substantiated in terms of statistical significance. This can be accomplished by making use of standard probability considerations. Let $M$ be a positive integer. For a large enough number of realizations $M$, we will search for an estimate of the mean $\langle P_{k}^{\delta}\rangle$ and the variance $\sigma_{P_{k}}^{2}$. Since the results represented in Fig. \ref{GOE106} have been calculated with a ensemble of $M=10^{6}$ spectra, they qualify as a good starting point. Let $P_{k}^{\delta}$ be a random variable with mean $\langle{P_{k}^{\delta}}\rangle$ and variance $\sigma^{2}_{P_{k}^{\delta}}$. By the well-known central limit theorem, the probability distribution of the mean estimator random variable $\langle P_{k}^{\delta}\rangle_{M}\equiv \frac{1}{M}\sum_{i=1}^{M}P_{k}^{(i)}$ is found to be \begin{equation}\label{distribution}
\langle P_{k}^{\delta}\rangle_{M}\sim\mathcal{N}\left(\langle P_{k}^{\delta}\rangle_{\infty},\frac{\sigma^{}_{P_{k}^{\delta}}}{\sqrt{M}}\right),
\end{equation}
where $\langle P_{k}^{\delta}\rangle_{\infty}\equiv \mu_{M}$ corresponds to the mean value for the power spectrum calculated with the ensemble of $10^{6}$ spectra, $\sigma_{P_{k}^{\delta}}\equiv\sqrt{\sigma_{P_{k}^{\delta}}^{2}}$ is now the standard deviation associated with it, $\sigma_{M}\equiv\sigma_{P_{k}^{\delta}}/\sqrt{M}$, and $1\ll M\lesssim10^{6}$ is an arbitrary number of realizations. The next step is therefore to investigate this new distribution as a function of $M$. It follows from Eq. \eqref{distribution} that the probability density function for the random variable $\langle P_{k}^{\delta}\rangle_{M}$ is simply given by the two-parameter expression
\begin{equation}\label{density} \rho_{M}(x)=\frac{1}{\sigma_{M}\sqrt{2\pi}}\exp\left[-\frac{(x-\mu_{M})^{2}}{2\sigma_{M}^{2}}\right].
\end{equation}
Note, however, that $\mu_{M}\neq \mu_{M}(M)$, since this value is fixed as a constant regardless of the value of $M$. 

To assess how incompatible the numerical data are with the theoretical formula for the GOE case we will immediately implement the $p$-values. We will assume the theoretical formula to be incorrect whenever $p<0.05$, that is, we will consider the adjustment to the theoretical curve to be due to chance. Otherwise, whenever $p\geq0.05$, the agreement will be taken statistically substantiated.

\begin{figure}[h]
\centering
\includegraphics[width=0.52\textwidth]{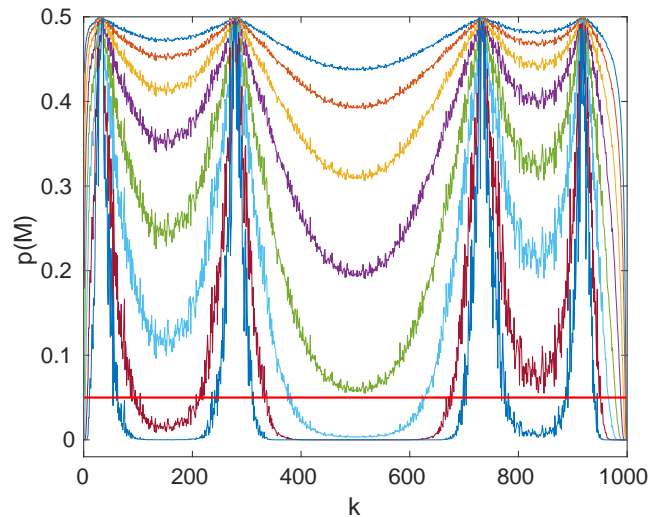}
\caption{(Color online) Values of $p(M)$ for, from top to bottom, $M=10,30,100,300, 1000,3000,10000,30000$. The value $p=0.05$ has been represented (red solid line) as the upper limit for which the theoretical curve remains invalid.}
\label{pvalues}
\end{figure}

\begin{figure}[h]
\centering
\includegraphics[width=0.52\textwidth]{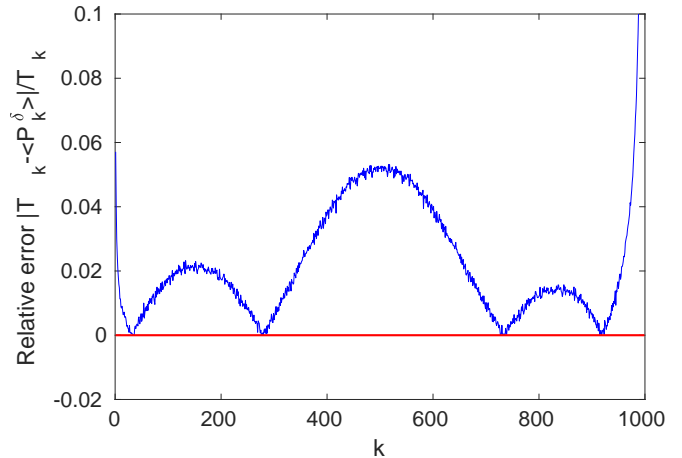}
\caption{(Color online) Relative error with respect to the theoretical values for the case in which $M=10^{6}$.}
\label{relativerror}
\end{figure}

Let $T_{k}$ denote de theoretical value for $\langle P_{k}^{\delta}\rangle_{\infty}$ given by Eq. \eqref{powergoe}. The $p$-value for each value of $M$, $p(k;M)$, is then taken to be
%
%\begin{widetext}
\begin{equation}\label{eqpvalue}
    p(k;M)\equiv\begin{cases}
{\displaystyle \int_{T_{k}}^{\infty}\mathrm{d}x\,\rho_{M}(x)},\,\,\,&T_{k}\geq\langle P_{k}^{\delta}\rangle_{\infty},\\[0.3cm]
{\displaystyle \int_{-\infty}^{T_{k}}\mathrm{d}x\,\rho_{M}(x)},\,\,\,&T_{k}\leq\langle P_{k}^{\delta}\rangle_{\infty}.\\
\end{cases}
\end{equation}
%\end{widetext}
%
%
%\begin{widetext}
%\begin{equation}\label{eqpvalue}
%    p-\textrm{value}(M)\equiv p(k;M)=\begin{cases}
%{\displaystyle \int_{T_{k}}^{\infty}\mathrm{d}x\rho_{M}(x)=\frac{1+\erf\left(\frac{\mu_{M}-T_{k}}{\sigma_{M}\sqrt{2}}\right)}{2}},\,\,\,&T_{k}\geq\langle P_{k}^{\delta}\rangle_{\infty},\\[0.3cm]
%{\displaystyle \int_{-\infty}^{T_{k}}\mathrm{d}x\rho_{M}(x)=\frac{1-\erf\left(\frac{\mu_{M}-T_{k}}{\sigma_{M}\sqrt{2}}\right)}{2}},\,\,\,&T_{k}\leq\langle P_{k}^{\delta}\rangle_{\infty},\\
%\end{cases}
%\end{equation}
%\end{widetext}
Since for each case the result depends strongly on $M$, this can be used to study the function $p(k;M)$ for certain values of the number of realizations. This is plotted in Fig. \ref{pvalues} for, from top to bottom, $M=10,30,100,300, 1000,3000,10000,30000$. The value $p=0.05$ has been represented as the upper limit for which the theoretical curve remains invalid.
%Note $p(k;M)$ is symmetric with respect to the axis $k=N/2=500$.
To better understand this result, we have also plotted the values of the relative error $|T_{k}-\langle P_{k}^{\delta}\rangle_{10^{6}}|/T_{k}$ as a function of the frequency in Fig. \ref{relativerror}. This is the relative error with respect to the expected value for the ensemble of $M=10^{6}$ spectra that we chose to establish an ensemble estimator via the central limit theorem, which means it will be representative for $1\ll M\lesssim 10^{6}$ as well. 

We conclude that for numbers of realizations $M\lesssim 10^{3}$, for all frequencies the inequality $p(M)>0.05$ is verified and, as a consequence, in all these cases it is reliable to take the agreement between theoretical expression and numerical data as statistically significant. However, for values $M\gtrsim10^{3}$ this is not guaranteed, and we observe indeed that $p(M)$ tends to zero for these cases in the same way the relative error increases for certain frequencies. For large enough values of $M$, the theoretical expression becomes valid only for a few points corresponding to the minima of the relative error. In particular, it is \textit{only} in the large-$M$ limit that the result is completely unacceptable for all frequencies, as exemplified by 
\begin{equation}\lim_{M\to\infty}p(k;M)=0,\,\,\forall k\in\{1,2,\dots,N-1\}.\end{equation} Also worth remarking is the observation that the maximum error occurs for the Nyquist frequency $\omega_{k_{\textrm{Ny}}}\equiv\pi$, i.e., $k_{\textrm{Ny}}\equiv N/2$.

This implies that the theoretical result derived in \cite{demo}, even though not fully correct, can be viewed as an extraordinary tool to analyze the chaoticity of a quantum system, so long as the realizations are not too many. Our results indicate that the corresponding expressions can thus be safely used. It is obvious that an explicit, easily applicable, closed-form expression for GOE is desirable, but it would not yield further significant improvements in spectral analyses. Indeed, the circumstances under which these results could be important constitute quite a small set of conditions. This manifests itself blatantly when one realizes that the general picture is that in which one may reasonably have only one set of eigenlevels with a relatively small size when compared with the dimension scales considered in this work.

With all the previous conclusions available to us, we propose one procedure to correctly perform any power spectrum analysis in Tab. \ref{protocol}.

\begin{table}
\begin{center}
\begin{tabular}{ l p{7cm} } 
\multicolumn{2}{c}{\textbf{Protocol for spectral statistical analysis}} \\
 \hline
 \textbf{Step 1.} & Unfold each spectrum individually, either analytically or numerically. If a complete spectrum with $N$ levels is divided into $M$ spectra with $N/M$ levels each, then unfold each spectrum of length $N/M$ separately too.  \\
 \textbf{Step 2.} & Proceed with the spectral analysis, and compare the results it yields with the expected expressions in \cite{demo}. \\
 \hline
\end{tabular}
\end{center}
\caption{Suggested protocol to follow for the unfolding procedure so that the analyzed results can be compared with the theoretical expressions for the power spectrum of quantum integrable and chaotic systems in \cite{demo}.}
\label{protocol}
\end{table}

\section{Application to real physical systems}
\label{section5}

\subsection{Characterization of the transition from integrability to chaos}

In this section we apply the previous protocol to a physical system that transits from the chaotic regularity class to the integrable one. 

The model under consideration is a Heisenberg spin-1/2 chain. Such a model has been proved to transit from integrability to chaos using the nearest neighbor spacing distribution \cite{lfsantos} as well as the the ratio of consecutive level spacings distribution \cite{transitionratios}. Furthermore, its importance transcends the discipline of quantum chaos. It constitutes a model for the many-body localization transition \cite{Pal10}. In \cite{Serbyn16} it has been shown that this fact implies a two-stage crossover in short-range spectral fluctuations. And in \cite{Buijsman19}, relying again on short-range spectral statistics, the Gaussian $\beta-$ensemble \cite{betaensemble} has been proposed as a model for this transition. Here, we will concentrate on the transition as seen by the $\delta_{n}$ statistic. 

The Hamiltonian of the Heisenberg spin-1/2 chain model is given by 
\begin{equation}\label{hamiltonian}
H=\sum_{n=1}^{L}\omega_{n}S_{n}^{z}+\varepsilon S_{d}^{z}+\sum_{n=1}^{L-1}J\hat{S}_{n}\cdot \hat{S}_{n+1},
\end{equation}
where $L$ is the number of sites, and $\hat{S}_{n}\equiv \vec{\sigma}_{n}/2$ are the spin operators located at site $n$ with $\vec{\sigma}_{n}$ being the Pauli spin matrices at that exact site. We briefly comment on the meaning of each of the terms in Eq. \eqref{hamiltonian}. The first term describes the Zeeman splitting of each spin $n$ as a consequence of the static magnetic field in the $z$-direction. In general, each $\omega_{n}$ is a random variable belonging to certain probability distribution. A site $d$ is called a defect if all the other sites are assumed to have the same energy splitting $\omega_{n}=\omega$ except the site $d$, where the splitting is $\omega+\varepsilon$ instead. Finally, two possible couplings between the nearest neighbor spins are described by the last term of Eq. \eqref{hamiltonian}. The first one is simply the diagonal Ising interaction, while the second is the off-diagonal flip-flop term, which is responsible for excitation propagation in the chain. Also, the chain is taken to be isotropic since $J$ is a constant that quantifies both the coupling strength between the Ising interaction and that of flip-flop term. Only two-body interactions can take place in this chain. 

For our simulation, the defect energy difference has been chosen to
vanish, $\varepsilon=0$, and the static magnetic field is random. We
have taken $J=1$, the number of sites $L=14$, we have imposed periodic boundary conditions, $\hat{S}_{15}=\hat{S}_1$, and we consider $\hbar=1$. In other words, we
will study the transition given rise by the one-parameter Hamiltonian
\begin{equation}\label{ourhamiltonian}
\mathcal{H}\equiv H(\omega_{n},\varepsilon=0)=\sum_{n=1}^{14}\omega_{n}S_{n}^{z}+\sum_{n=1}^{14}\hat{S}_{n}\cdot \hat{S}_{n+1}.
\end{equation}
Each realization consists of $L=14$ random numbers, $\omega_{n}$, whose probability
distribution is uniform in the range $\left[-\omega, \omega \right]$. Since $[H,\hat{S}_{z}]=0$, we consider
the sector with $\hat{S}_{z}=0$. In this case, the dimension of the
Hilbert space is $d=\mqty(14 \\ 7)=3432$. However, we use the central part of the spectrum only, since it gives a better description of the many-body localization crossover than the complete spectrum \cite{Serbyn16,Buijsman19}. Hence, the size of each spectrum
has been therefore chosen $N=1144$ ($1/3$ of the total number of energy levels), and we have simulated $M=1000$
realizations. Besides this, we have excluded the first and last $30$ levels both
before and after the unfolding process, in order to prevent spurious
effects from happening in our analysis. The polynomial to which we
have fitted the original set of energies $\{E_{i}\}_{i=1}^{N}$ to
obtain the unfolded sequence $\{\epsilon_{i}\}_{i=1}^{N}$ was taken to
be of degree $6$, which means the smooth part of the cumulative density
function was of the form $\overline{N}(x)=\sum_{k=0}^{6}a_{k}x^{k}$,
with coefficients $\{a_{k}\}_{k=0}^{6}$ that resulted from the best
possible fit to the sequence $\{E_{i}\}_{i=1}^{N}$. Then, it is
straightforward to evaluate
$\eval{\overline{N}(x)}_{x=E_{i}}=\epsilon_{i}$, $\forall
i\in\{1,2,\dots,N\}$. 

\begin{center}
\begin{figure*}
\begin{tabular}{ll}
%\hspace{-2cm} & 
\includegraphics[width=1.15\textwidth, trim={2cm 0 7cm 0},clip, angle=-90]{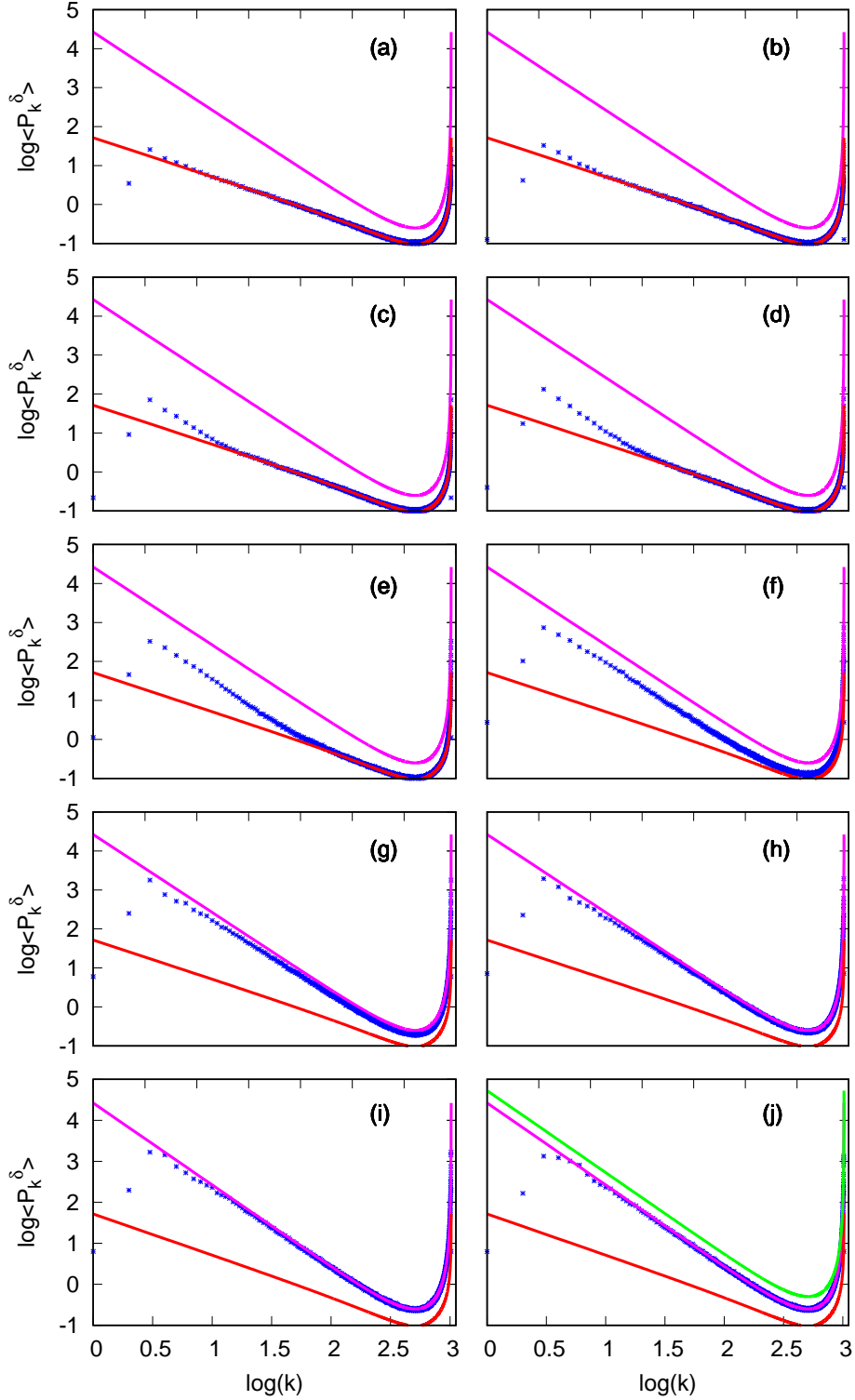}
\end{tabular}
\caption{(Color online) Power spectrum transition of the system with eigenvalues obtained from $\mathcal{H}$, Eq. \eqref{ourhamiltonian}, for the parameter values $\omega\in\{0.4,0.6,0.8,1,1.4,2,3,4,5,7\}$  from panels $(a)$ to $(j)$ in ascending order (blue points). Plotted against them are the theoretical power spectrum curve for GOE given in Eq. \eqref{powergoe} (red, solid line) and the leading order of the theoretical power spectrum for GDE given in Eq. \eqref{power4} (magenta, solid line). Additionally, Eq. \eqref{power2} has been represented in panel $(j)$ (green solid line) for comparison. An average over $M=1000$ spectra with $N=1144$ levels each has been performed. }
\label{transition}
\end{figure*}
\end{center}

In Fig. \ref{transition} we show the results for the power spectrum of the $\delta_n$ statistics for the noise parameter
values $\omega_{}\in\{0.4, 0.6, 0.8, 1.0, 1.4, 2.0, 3.0, 4.0, 5.0, 7.0\}$. The many-body localization transition is believed to occur around $\omega_c \sim 3.6$ \cite{Luitz15}. Hence, chaotic-like spectral fluctuations are expected for $\omega \ll \omega_c$, and integrable ones for $\omega \gg \omega_c$. From the $10$ panels of Fig. \ref{transition} (see caption for more details) we highlight the following conclusions:

{\em (i)} The system reaches the ergodic limit at very low values of $\omega$. In panel (a), corresponding to $\omega=0.4$, we can see that the power spectrum coincides with the theoretical results for GOE if $k \gtrsim 3$. Lower frequencies display nonuniversal features, probably due to the unfolding procedure \cite{misleadingsign}. 

{\em (ii)} In the range $\omega \in  [0.6,1.4]$, corresponding to panels from (b) to (e), we observe a mixture of behaviors. In all the cases, the numerical and the GOE theoretical curves coincide in the high-frequency region. However, below a certain {\em critical} frequency, that we denote $k_c$, the slope changes abruptly, and the numerical set of points approximately display $1/f^2$ noise, trademark of integrable systems. In panel (b), corresponding to $\omega=0.6$, $k_c \sim 10$. This frequency represents a distance between energy levels around $\Delta\epsilon \sim 100$. In other words, when $\omega=0.6$ the correlations between energy levels are GOE-like if there are less than other $100$ levels between them, and become Poisson-like in the opposite case. In panel (e), showing the case $\omega=1.4$, the critical frequency is $k_c \sim 100$, corresponding to a distance between energy levels $\Delta\epsilon \sim 10$.

{\em (iii)} For $2 \lesssim \omega \lesssim 4$, we see a fast transition at high frequencies, corresponding to short-range correlations. Panel (f), corresponding to $\omega=2$, shows that the shortest-range correlations are still close to GOE, whereas the largest ones are however close to Poisson. On the contrary, panel (h), corresponding to $\omega=4$, shows the opposite results: shortest-range correlations are compatible with a Poisson-like behavior, where we can still see a small deviation at $k \lesssim 10$ (as we have pointed out above, frequencies below $k \sim 3$ display nonuniversal effects, probably due to the unfolding procedure).

{\em (iv)} Finally, the crossover to regularity seems totally completed at $\omega \sim 7$, case displayed in panel (j).

From these results we can conclude that the original theoretical results for the power spectrum \cite{demo} provide a very good and complete picture. Indeed, we also display in panel (j) of Fig. \ref{transition} the result obtained in \cite{losenemigos} for the integrable limit. We clearly see that this limit is never reached. Hence, if one were to make use of such an expression to study the exact same transition an intermediate conclusion, belonging \textit{neither} in the integrable \textit{nor} in the chaotic case, would be reached. This would be in clear contradiction with the dynamics transition that is however well-established for this system, thus resulting in an blatantly wrongful verdict. 

\subsection{Phenomena only accessible to long-range statistics}

After showing that the equations derived in \cite{demo} can be safely used to characterize the regularity of real complex many-body quantum systems, we profit from this fact to delve into the nature of the many-body localization crossover. To do so, we compare the results of Fig. \ref{transition} with short-range spectral fluctuations analyzed by means of the ratio of consecutive levels, used in \cite{Buijsman19} to propose that the Gaussian $\beta$-ensemble provides a good description of the crossover.

\begin{figure}[h]
\hspace*{-1.1cm}
\includegraphics[width=0.40\textwidth, angle=-90]{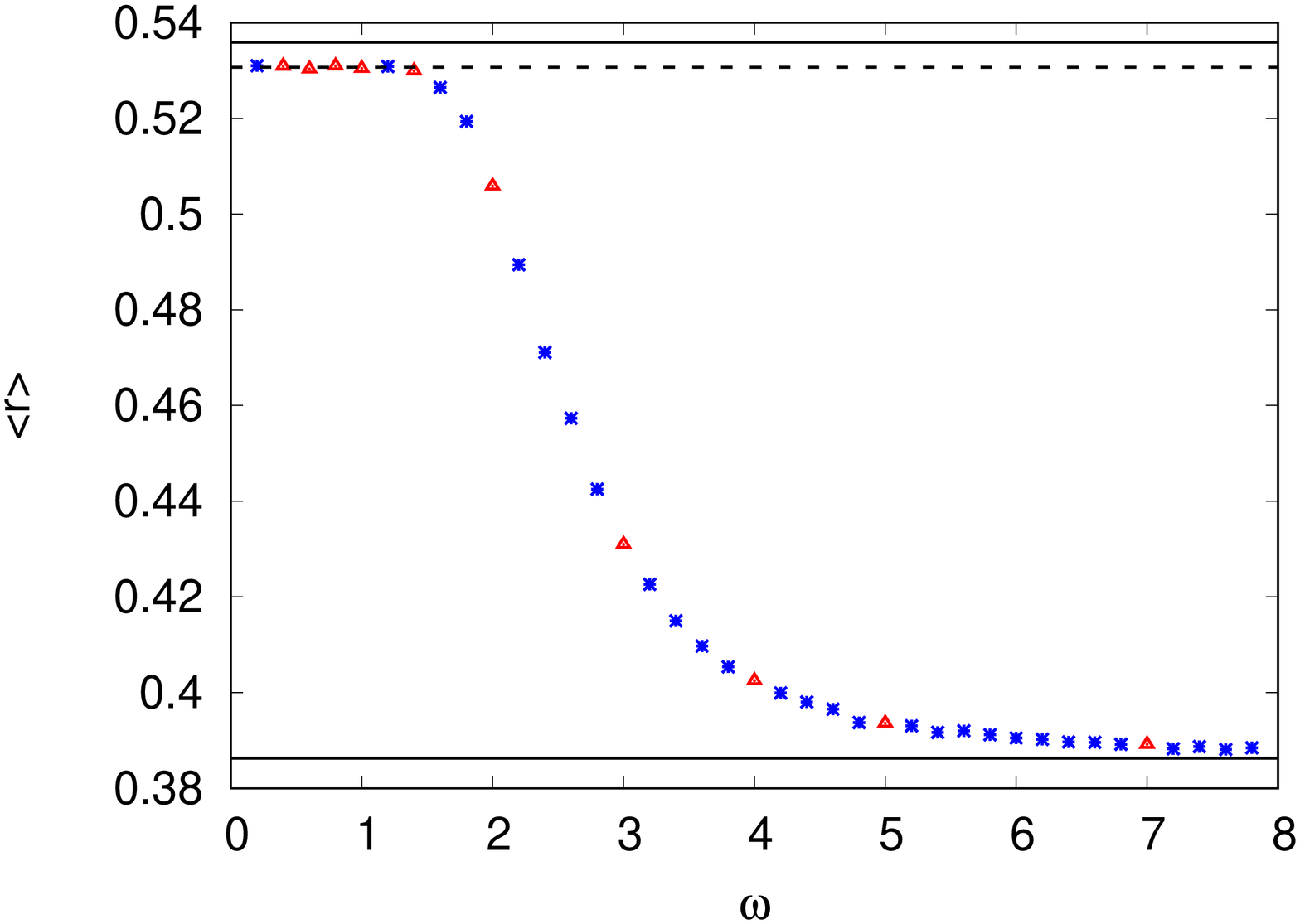}
\caption{(Color online) Mean value of the ratios Eq. \eqref{minratios} as a function of the transition inducing parameter $\omega$ performed with the $N=1144$ eigenlevels of the Heisenberg spin$-1/2$ chain, Eq. \eqref{ourhamiltonian}, with a number of $M=1000$ realizations each (blue points). The averages corresponding to the same values of $\omega$ as the power spectra shown in Fig. \ref{transition} are plotted for clarity (red triangles). The integrable limit, $\left<\widetilde{r}_{\text{Poisson}} \right> = 2\ln2-1\approx 0.38629$ and the GOE one, $\left<\widetilde{r}_{\text{GOE}} \right> = 4-2\sqrt{3}\approx 0.53590$, have been plotted too (black, lower and upper solid lines, respectively). For comparison, the mean ratio corresponding to our generated GOE of dimension $N=10^{3}$ and $M=10^{6}$ realizations, $\left<\widetilde{r}_{\text{M}} \right> \approx 0.5307$, has been shown too (black, dashed line).}
\label{meanvalue}
\end{figure}

The ratio of consecutive level spacings statistic, $r$, is defined as the random variable taking on values
\begin{equation}\label{ratio}
r_{n}\equiv \frac{s_{n}}{s_{n-1}},\,\,\textrm{where}\,\,s_{n}\equiv E_{n+1}-E_{n},\,\,\forall n\in\{2,\dots,N-1\},
\end{equation}  
and $\{E_{n}\}_{n=1}^{N}$ is a complete set of energies in ascending order, that is, verifying $E_{n}\geq E_{m}$ whenever $n\geq m$, coming from a quantum system spectrum. Note that the construction of this quantity does not involve the unfolding procedure in any way, which clearly differs from what the situation is for the $\delta_{n}$ or the NNSD. When no theoretical distribution is employed, quantum chaoticity can be analogously characterized by the mean value $\langle \widetilde{r}\rangle$, where now 
\begin{equation}\label{minratios}
    \widetilde{r}_{n}\equiv \frac{\min(s_{n},s_{n-1})}{\max(s_{n},s_{n-1})}=\min\left(r_{n},\frac{1}{r_{n}}\right),
\end{equation}
which always exists. This modification is equivalent to studying the ratios distribution over the support $[0,1]$. Theoretical values of $\langle \widetilde{r}\rangle$ for the integrable as well as the GOE chaotic case have been derived from $3\times 3$ random matrices in a Wigner-like spirit and yield, respectively, $\left<\widetilde{r}_{\text{Poisson}} \right> = 2\ln2-1$ and  $\left<\widetilde{r}_{\text{GOE}} \right> = 4-2\sqrt{3}$. The eigenlevels yielded by the diagonalization of $\mathcal{H}$, Eq. \eqref{ourhamiltonian}, allow us to obtain the ratios Eq. \eqref{ratio}, from which its expression in the form of Eq. \eqref{minratios} follows directly. We then perform an average over the total number of ratios for each value of the transition inducing parameter $\omega$, and end up with the mean estimators $\langle \widetilde{r}\rangle\equiv\langle \widetilde{r}\rangle(\omega)$. These are shown in Fig. \ref{meanvalue} as a function of $\omega$. We also plot both limiting integrable and GOE cases with a black straight line. And we also include the mean ratio numerically obtained with our data of $M=10^{6}$ GOE matrices of dimension $N=10^{3}$, plotted with dashed line. This result differs from the theoretical prediction for GOE matrices, because it is derived for $3\times 3$, and small, but maybe  significant differences, are expected for larger matrices \cite{ratios}.

Results shown in Fig. \ref{meanvalue} play a very valuable role in this part of our work because they quite manifestly prove that there is a fundamental difference between measurements given by long-range and short-range spectral statistics, here exemplified by the $\delta_{n}$ and the $P(r)$, respectively. In Fig. \ref{meanvalue} we see a fast crossover between $\omega \sim 1.8$ and $\omega \sim 4$, but no traces of the deviation from a GOE-like behavior shown in Fig \ref{transition} for $\omega \lesssim 1.4$, and no clues about the consequences of the crossover on different correlation scales.  This allows us to obtain statistically significant comparisons between our model and the mean ratios value, because the amount of available data is now similar. One can check that, then, the distribution of the ratios detects a \textit{completely} chaotic behavior for $\omega\lesssim1.4$, whereas already for $\omega=1.4$ the $\delta_{n}$ power spectrum shows a clear separation between the simulated and theoretical curves.  This strongly suggests that ergodicity is a much more strict condition for the $\delta_{n}$ power spectrum, meaning that this statistic is considerably more sensitive to it than short-range ones, like the distribution of the ratios, are. Hence, to get a safe conclusion about whether a quantum system is fully ergodic or not, the analysis of long-range spectral statistics, like the $\delta_n$ power spectrum is mandatory. Even though it is true that short-range statistics have been widely used to get a first idea of what the regularity class of a quantum system might be, a more involved investigation needs to take into account both short-range \textit{and} long-range outcomes.

It is very recently that one particular model has been claimed to offer a universal description of the spectral fluctuations along the many-body localization crossover \cite{Buijsman19}. Also known as the Continuous Gaussian Ensemble, the $\beta-$ensemble is a generalization of the classical Gaussian ensembles which was in its origins studied as a theoretical joint eigenvalue distribution with applications, for instance, in lattice gas theory \cite{lattgas2}. This eigenvalue distribution can be derived from an ensemble of random matrices \cite{betaensemble}. The Gaussian $\beta-$ensemble has since been used for various purposes \cite{LeCaer07,powerspectrum}. This ensemble essentially consists of tridiagonal, real, and symmetric matrices whose entries are classical random variables, these being normal, $\mathcal{N}(\mu,\sigma)$ with $\mu$ being its mean and $\sigma\equiv\sqrt{\sigma^{2}}$ its standard deviation, and chi, $\chi_{k}\equiv \sqrt{\chi_{k}^{2}}$ with $k\in\mathbb{R}_{+}\cup\{0\}$ denoting a continuous, non-negative number of degrees of freedom. The matrix elements of the model $\mathcal{H}_{i,j}\equiv(\mathcal{H})_{i,j}$ are random variables distributed over $\mathbb{R}$ with distribution given by
\begin{equation}
    \mathcal{H}_{ii}\sim\mathcal{N}\left(0,\sqrt{\frac{1}{2\lambda}}\right), \,\,\,\forall i\in\{1,2,\dots,N\},
    \end{equation}
    and
    \begin{equation}
    \mathcal{H}_{i+1,i}=\mathcal{H}_{i,i+1}\sim\sqrt{\frac{1}{4\lambda}}\chi_{(N-i+1)\beta},\,\,\,\forall i\in\{1,2,\dots,N-1\},
\end{equation}
with $\lambda,\beta\in\mathbb{R}_{+}$ being free parameters. The values $\beta=0,1,2,4$ correspond to Poisson, GOE, GUE, and GSE, respectively, but the continuous nature of $\beta$ generalizes the pattern of thinking that the RME classification is discrete, and thus can take any real, positive value. The convention that $\chi_{0}\equiv 0$ is assumed.

In \cite{Buijsman19}, agreement with the level statistics of the physical system Eq. \eqref{ourhamiltonian} over the entire crossover range from thermal to the many-body localized phase was found. All results and conclusions are calculated with and referred to the distribution of the ratio of consecutive level spacings, $P(r)$, and comparisons are offered making use of the statistic given in Eq. \eqref{minratios} together with the $\beta-$ensemble. Here we apply the same logic and study whether this matrix ensemble can or cannot correctly describe the integrable-GOE crossover for the $\delta_{n}$ power spectrum. 

\begin{center}
\begin{figure*}
\begin{tabular}{ll}
\hspace{1.1cm} & 
\includegraphics[width=0.8\textwidth, trim={2cm 0 3cm 0},clip, angle=-90]{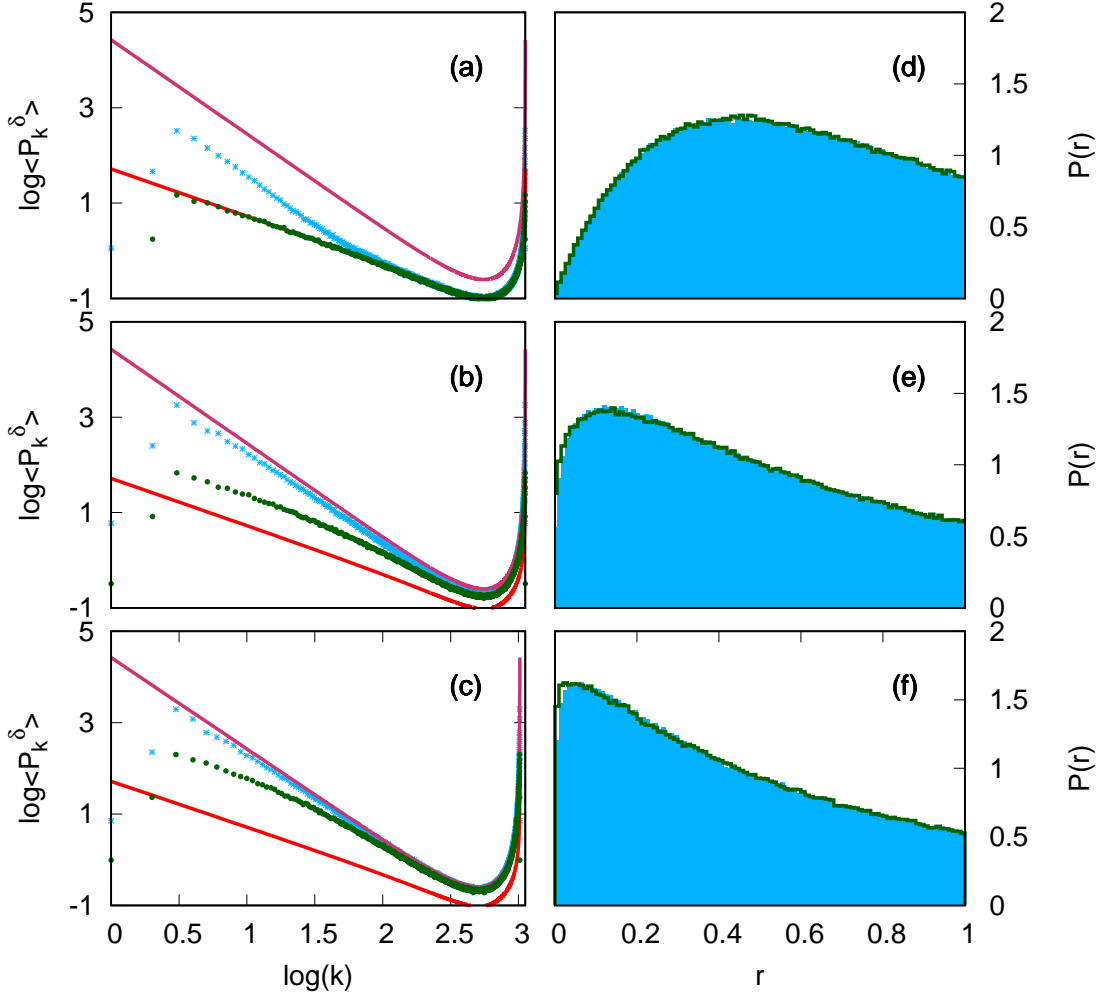}
\end{tabular}
\caption{(Color online) Power spectrum transition of the system with eigenvalues obtained from $\mathcal{H}$, Eq. \eqref{ourhamiltonian}, for the parameter values $\omega\in\{1.4,3,4\}$ (blue, $+$ symbols), together with the power spectrum for the same $\omega$ obtained by means of the $\beta-$ensemble with $\beta\in\{0.99398,0.21067,0.06732\}$ (green points), panels $(a)-(c)$ in ascending order. Plotted against them are the theoretical power spectrum curve for GOE given in Eq. \eqref{powergoe} (red, solid line) and the leading order of the theoretical power spectrum for GDE given in Eq. \eqref{power4} (magenta, solid line). Distribution of the ratio of consecutive level spacings Eq. \eqref{minratios} coming from the system Eq. \eqref{ourhamiltonian} for the parameter values $\omega\in\{1.4,3,4\}$ (blue, filled histogram), together with the ratios distribution for the same $\omega$ obtained by means of the $\beta-$ensemble with $\beta\in\{0.99398,0.21067,0.06732\}$ (green, solid lines), panels $(d)-(f)$ in ascending order. An average over $M=1000$ spectra with $N=1144$ levels each has been performed.}
\label{powerratios}
\end{figure*}
\end{center}

We proceed as follows. First, we take the numerically obtained values $\langle \widetilde{r}\rangle$ and $\beta$ depicted in Table I of \cite{Buijsman19}, and perform a polynomial fit of degree 4, which provides a function $\beta=\beta(\langle \widetilde{r}\rangle)$. Next, we take the mean values $\langle \widetilde{r}\rangle$ shown in Fig. \ref{meanvalue}, and make use of the former expression to find the corresponding values of the $\beta$ parameter. With these in our possession, we are then able to simulate the Gaussian $\beta-$ensemble for each value of $\beta$ recently calculated, and we do so a number of $M=1000$ times following precisely the same procedure explained in \cite{Buijsman19}, obtaining a number of $N=1144$ eigenlevels per realization that follow from diagonalization. We are now naturally capable of constructing the $\delta_{n}$ power spectrum for this set of levels as well as the histogram that gives the distribution $P(r)$ for each value of the transition parameter $\omega$. These two important quantities are shown in Fig. \ref{powerratios} for $\omega\in\{1.4,3,4\}$. We observe a near-perfect match between the distribution of the ratios of our Hamiltonian and that coming from the simulation of the $\beta-$ensemble for the corresponding value of $\beta$, meaning $P(r)$ can be indeed correctly characterized by this particular ensemble of matrices. High-lying frequencies in the power spectrum, which correspond with short-range correlations, do indeed corroborate this fact, for the agreement between the corresponding two curves is, again, remarkable. Notwithstanding, representations of the power spectrum for low-lying frequencies have nothing in common with each other and show significant deviation. This is due once more to the extreme sensitivity of long-range spectral statistics, and hints at the impossibility to universally describe quantum crossovers simply by means of the $\beta-$ensemble, since this is \textit{only} true for short-range statistics such as the $P(r)$ and possibly some others. As far as the $\delta_{n}$ statistic goes, and by extension a whole class of long-range statistics, this has been proven to no longer be the case. 

%It is noteworthy to observe that the power spectrum for the $\beta-$ensemble \textit{is} completely ergodic for $\beta=0.99398$, which means that the deviation at low-lying frequencies with respect to the theoretical curve shown by the spin chain power spectrum for $\omega=1$ \textit{cannot be justified with the unfolding procedure}, for it has been applied to \textit{both} spectra to construct the $\delta_{n}$ statistic, implying that this feature is physical. 

\section{Conclusions}
\label{section6}

The study of quantum many-body regularity class can be undertaken by means of spectral statistics. One of such statistics frequently used in the analysis is the $\delta_{n}$, whose power spectrum is of interest because it relates the degree of chaoticity of a system with an universal noise of the form $1/f$ and $1/f^{2}$ for chaotic and integrable systems, respectively. Theoretical representations for the expected values of the power spectra have been derived in the past. Our interest has been focused on the classical ensembles that describe the most frequent physical systems, that is, Poisson and GOE. 

In this work we have established the role that the process of unfolding plays in spectral analysis, namely, that it introduces spurious correlations. However, since level repulsion for quantum chaotic dynamics means in these systems correlations already exist, this is of utter importance for quantum integrable systems only, since in principle correlations should not be present in their characterization. A consequence that cannot go unnoticed is that, as a result, the GDE power spectrum obtained by the usual procedure of unfolding deviates from its expected theoretical expression by a factor of $2$, which makes itself evident in the analysis. Quite surprisingly, this factor entails that the original results published in \cite{demo} are correct. This effect cannot be escaped, \textit{not even by proceeding with a perfect, exact, analytic unfolding transformation}. In this work, we have therefore provided a protocol to follow if one wishes to succeed in comparing quantum integrable power spectra with theoretical values. 

The GOE case is doubtlessly less fortunate, since we are still lacking an exact representation for its theoretical power spectrum expected value, which needs to be found by considering correlations to all orders. This means the approximate form derived in \cite{demo} is currently our best possible alternative. By performing a stringent, rigorous statistical test on this result, we have proved that this can be regarded as an excellent tool to guarantee that a quantum system belongs in the chaotic class, and can be safely made use of in a wide range of situations that are often the case. 

We have displayed an explicit transition of a system from the chaotic regime to the integrable one by means of the protocol we previously introduced, and conclude that it does indeed afford excellent results. We have chosen, for this purpose, a Heisenberg $1/2$-spin chain, which accounts for the crossover from the thermal to the many-body localized phase.

Our work concludes with a study of the differences in the determination of quantum chaoticity when using long-range statistics in contrast with short-range ones, like the distribution of the ratios of consecutive level spacings. We find that consequences of crossovers on different correlation scales cannot be afforded by short-range spectral statistics and that the Gaussian $\beta-$ensemble fails to describe long-range correlations, which are completely inaccessible for short-range statistics, for which predictions from this ensemble are valid only. A richer, full characterization of quantum chaoticity must, therefore, necessarily consider information of both long and short-range results.  

~

\begin{acknowledgements}

This work has been supported by the Spanish Grants
Nos. FIS2015-63770-P (MINECO/ FEDER) and PGC2018-094180-B-I00 (MCIU/AEI/FEDER, EU). The authors are grateful to L. Mu\~{n}oz for enlightening discussions.
\end{acknowledgements}

\begin{widetext}
\appendix*
\section{Calculation of the theoretical power spectrum for the integrable case}
In this section we give details on the calculations performed in Section III, for both the unfolded and the reunfolded cases. 
\subsection{Derivation for the unfolded spacings}
In evaluating Eq. \eqref{pk} the first step is to find the value of $\langle \delta_{\ell}\delta_{m}\rangle$. One expands this term to obtain 
\begin{equation}\label{a1}
\langle \delta_{\ell}\delta_{m}\rangle=\sum_{i=1}^{\ell}\sum_{j=1}^{m}\langle (s_{i}-1)(s_{j}-1)\rangle=\sum_{i=1}^{\ell}\sum_{j=1}^{m}(\langle s_{i}s_{j}\rangle -\langle s_{i}\rangle - \langle s_{j}\rangle+1). 
\end{equation}
Concentrating on each of the terms separately yields 
\begin{equation}\label{a2}
    \sum_{i=1}^{\ell}\sum_{j=1}^{m}\langle s_{i}s_{j}\rangle=\sum_{i=1}^{\ell}\sum_{j=1}^{m}(\delta_{ij}+1)=\min\{\ell,m\}+\ell m
\end{equation}
as well as 
\begin{equation}\label{a3}
    \sum_{i=1}^{\ell}\sum_{j=1}^{m}\langle s_{i}\rangle=\sum_{i=1}^{\ell}\sum_{j=1}^{m}1=\ell m=\sum_{i=1}^{\ell}\sum_{j=1}^{m}\langle s_{j}\rangle.
\end{equation}
Plugging Eqs. (\ref{a2}) and (\ref{a3}) into Eq. \eqref{a1} one immediately obtains for the power spectrum the expression in Eq. \eqref{pk2}. Summing over $\ell$ and $m$ leads to 
\begin{equation}\label{a4}
\langle P_{k}^{\delta}\rangle=\frac{1}{\sin^{2}(\omega_{k}/2)}\frac{1+2N-\cos(\omega_{k}N)+\frac{\sin (\omega_{k}N)}{\tan(\omega_{k}/2)}}{4N},
\end{equation}
which is then rearranged to finally yield the result in Eq. \eqref{power2}. One must take the time to observe that such a leading order expression is indeed \textit{inexact} when one considers an arbitrary value of $N$. 
\subsection{Derivation for the reunfolded spacings}
We now consider Eq. \eqref{pcorrk}. The calculation that follows is analogous to the previous one, but caution must be exercised since it becomes more tedious. First one considers as starting point the equation 
\begin{equation}\label{a5}
\langle \widetilde{\delta}_{\ell}\widetilde{\delta}_{m}\rangle=\sum_{i=1}^{\ell}\sum_{j=1}^{m}\langle (\widetilde{s}_{i}-1)(\widetilde{s}_{j}-1)\rangle=\sum_{i=1}^{\ell}\sum_{j=1}^{m}(\langle \widetilde{s}_{i}\widetilde{s}_{j}\rangle -\langle \widetilde{s}_{i}\rangle - \langle \widetilde{s}_{j}\rangle+1). 
\end{equation}
We now take the easiest non-trivial element, i.e., 
\begin{equation}\label{a6}
\langle \widetilde{s}_{i}\rangle=N\left\langle\frac{s_{i}}{\sum_{j=1}^{N}s_{j}}\right\rangle.
\end{equation}
To proceed, we make use of the fact that the nearest neighbor spacings are uncorrelated in an integrable system. Suppose each random variable $\widetilde{s}_{i}$ is distributed as $\widetilde{s}_{i}\sim \textrm{Exp}(1)$, that is, $\widetilde{s}_{i}$ is a Poisson random variable with probability density $f_{i}$. One then has the $N$-dimensional integral for the mean $i$-th nearest neighbor reunfolded spacing given by
\begin{equation}
%\label{a7}
\begin{split}
\langle \widetilde{s}_{i}\rangle&=N\int_{\mathbb{R}^{N}}\mathrm{d}s_{1}\cdots \mathrm{d}s_{i}\cdots \mathrm{d}s_{N}\frac{s_{i}}{s_{1}+\dots+s_{i}+\dots+s_{N}}f_{i}(s_{1},\dots,s_{i},\dots,s_{N})
\\&=N\int_{0}^{\infty}\mathrm{d}s_{1}\cdots \mathrm{d}s_{i}\cdots \mathrm{d}s_{N}\frac{s_{i}}{s_{1}+\dots+s_{i}+\dots+s_{N}}e^{-(s_{1}+\dots+s_{i}+\dots+s_{N})}.
\end{split}
\end{equation}
This integral can be evaluated by chaning to hyperspherical coordinates. Performing the above integral produces 
\begin{equation}\label{a8}
\langle \widetilde{s}_{i}\rangle
%=N2^{N}I_{r}I_{\phi_{1}}\cdots I_{\phi_{N-1}}
=1.
\end{equation}
%where 
%\begin{equation}\label{a9}
%I_{r}\equiv \int_{0}^{\infty}\mathrm{d}r\, r^{2N-1}e^{-r^{2}}=\frac{(N-1)!}{2},
%\end{equation}
%\begin{equation}\label{a10}
%I_{\phi_{1}}\equiv\int_{0}^{\pi/2}\mathrm{d}\phi_{1}\,\sin^{2N-3}\phi_{1}\cos^{3}\phi_{1}=\frac{1}{2N(N-1)},
%\end{equation}
%and
%\begin{equation}\label{a11}
%I_{\phi_{i}}\equiv \int_{0}^{\pi/2}\mathrm{d}\phi_{i}\,\sin^{2(N-i)-1}\phi_{i}\cos\phi_{i}=\frac{1}{2(N-i)},\,\,\,i=2,\dots,N-1.
%\end{equation}
We observe that $\langle \widetilde{s}_{i}\rangle=\langle s_{i}\rangle$, $\forall i\in\{1,\dots, N\}$. This proves the re-unfolding process preserves the mean value of the nearest neighbor spacings. Clearly, $\langle s_{j}\rangle$ is calculated in the exact same manner and yields the same result. 

Next we consider the re-unfolded nearest neighbor spacing correlation given by 
\begin{equation}\label{a12}
\langle \widetilde{s}_{i}\widetilde{s}_{j}\rangle=N^{2}\left\langle\frac{s_{i}s_{j}}{\sum_{k=1}^{N}\sum_{\ell=1}^{N}s_{j}s_{\ell}}\right\rangle.
\end{equation}
Eq. \eqref{a12} requires considering two different cases: $(a)$ $i=j$ and $(b)$ $i\neq j$. For $(a)$ one has 
\begin{equation}\label{a13}
\langle \widetilde{s}_{i}^{2}\rangle=N^{2}\left\langle \frac{s_{i}^{2}}{\sum_{k=1}^{N}\sum_{\ell=1}^{N}s_{j}s_{\ell}}\right\rangle=N^{2}\int_{0}^{\infty}\mathrm{d}s_{1}\cdots \mathrm{d}s_{N}\frac{s_{i}^{2}}{(s_{1}+\dots+s_{N})^{2}}e^{-(s_{1}+\dots+s_{N})}=\frac{2N}{N+1},
\end{equation}
while for $(b)$ this becomes 
\begin{equation}\label{a14}
\langle \widetilde{s}_{i}\widetilde{s}_{j}\rangle=N^{2}\left\langle \frac{s_{i}s_{j}}{\sum_{k=1}^{N}\sum_{\ell=1}^{N}s_{j}s_{\ell}}\right\rangle=N^{2}\int_{0}^{\infty}\mathrm{d}s_{1}\cdots \mathrm{d}s_{N}\frac{s_{i}s_{j}}{(s_{1}+\dots+s_{N})^{2}}e^{-(s_{1}+\dots+s_{N})}=\frac{N}{N+1}.
\end{equation}
Both cases $(a)$ and $(b)$ can be expressed compactly by setting 
\begin{equation}\label{a15}
\langle \widetilde{s}_{i}\widetilde{s}_{j}\rangle=\frac{N(\delta_{ij}+1)}{N+1}.
\end{equation}
Note that Eq. \eqref{a15} is identical to Eq. \eqref{conditionb} in the limit $N\to\infty$. It is surprising that \textit{this difference is responsible for the $2$ factor that comes up in the power spectrum}. We shall now prove this. 

Plugging Eqs. (\ref{a6}) and (\ref{a15}) into Eq. \eqref{a5}, we obtain Eq. \eqref{corr4}:
\begin{equation}\label{a16}\begin{split}
\langle \widetilde{\delta}_{\ell}\widetilde{\delta}_{m}\rangle&=\sum_{i=1}^{\ell}\sum_{j=1}^{m}\left[\frac{N(1+\delta_{ij})}{1+N}-1\right]=\frac{N}{N+1}\left(\min\{\ell,m\}+\ell m\right)-\ell m\\ &=\frac{1}{N+1}\left(N\min\{\ell,m\}-\ell m\right).
\end{split}
\end{equation}
The power spectrum Eq. \eqref{pcorrk2} is therefore obtained straightforwardly inserting Eq. \eqref{a16} in Eq. \eqref{pcorrk}. To proceed, one must perform the two double sums that appear in it. The first one has already been calculated and yields the result Eq. \eqref{power2} with the substitution $N\to N+1$. The second sum is calculated to obtain 
\begin{equation}\label{a17}\begin{split}
    \frac{1}{N(N+1)}\sum_{\ell=1}^{N}\sum_{m=1}^{N}\ell m e^{i\omega_{k}(\ell-m)}&=\frac{e^{-i(N-1)\omega_{k}}}{N(N+1)(e^{i\omega_{k}}-1)^{4}}\left[N+e^{2i(N+1)\omega_{k}}-(N+1)e^{i\omega_{k}}-(N+1)e^{i(1+2N)\omega_{k}}\right.\\ &\left.-N(N+1)e^{i\omega_{k}}-N(N+1)e^{i(N+2)\omega_{k}}+2(N^{2}+N+1)e^{i(N+1)\omega_{k}}\right].
\end{split}
\end{equation}
On using the corresponding value for each sum, Eq. \eqref{power4} is eventually reached, after tedious but trivial algebra, as
\begin{equation}\label{a18}\begin{split}
\langle \widetilde{P}_{k}^{\delta}\rangle&=\frac{N}{4(N+1)\sin^{2}(\omega_{k}/2)}+\order{\frac{1}{N}}=\frac{1}{4\sin^{2}(\omega_{k}/2)}-\frac{1}{4(N+1)\sin^{2}(\omega_{k}/2)}+\order{\frac{1}{N}}\\&=\frac{1}{4\sin^{2}(\omega_{k}/2)}+\order{\frac{1}{N}}.
\end{split}
\end{equation}
It is worth noting that \textit{all} the calculations in this appendix are \textit{exact}. We conclude that the expressions usually used for the theoretical power spectrum in the integrable case are the leading order of the results here derived, and the expressions become exact as well when $N\to\infty$. 
\end{widetext}
\end{document}